\documentclass[conference]{IEEEtran}

\usepackage{amsmath} 
\usepackage{amsfonts} 
\usepackage{amssymb}
\usepackage{amsthm}
\usepackage[utf8]{inputenc}
\usepackage{subfigure}
\usepackage{caption}
\usepackage{bbm}

\ifCLASSINFOpdf 
\usepackage[pdftex]{graphicx} 
\usepackage[pdftex,colorlinks,
           bookmarks=true,%
            pdftitle=Modeling\ Static\ Cellular\ Cooperation\ by\ the\ Nearest\ Neighbour\ Model,%
           pdfauthor=A.\ Giovanidis\ -\ L.D.\ Alvarez Corrales -\ L.\ Decreusefond]{hyperref}  
\usepackage[numbers,sort&compress]{natbib} 
\usepackage{hypernat} 
\else

\usepackage[dvips]{graphicx} 
\usepackage[dvips,colorlinks,draft, 
           bookmarks=true,%
           pdftitle=Modeling\ Static\ Cellular\ Cooperation\ by\ the\ Nearest\ Neighbour\ Model,%
           pdfauthor=A.\ Giovanidis\ -\ L.D.\ Alvarez Corrales -\ L.\ Decreusefond]{hyperref} 
 \usepackage[numbers,sort&compress]{natbib} 
\fi 
\hypersetup{
   bookmarksnumbered,
   pdfstartview={FitH},
   citecolor={blue},
   linkcolor={red},
   urlcolor=[rgb]{0,0.55,0},
   pdfpagemode={UseOutlines}}

\DeclareMathOperator*{\argmin}{argmin}

\newtheorem{defi}{Definition}

\newtheorem{cor}{Corollary}
\newtheorem{theo}{Theorem}
\newtheorem{lem}{Lemma}
\newtheorem{NR}{Numerical Result}

\begin{document}

\title{Analyzing Interference from Static Cellular Cooperation using the Nearest Neighbour Model}

\author{Anastasios Giovanidis, Luis David \'Alvarez Corrales and Laurent Decreusefond\\
T\'el\'ecom ParisTech \& CNRS - LTCI, 23 avenue d'Italie, 75013 Paris, France\\
Contact: \{firstname.lastname\}@telecom-paristech.fr
}

\maketitle

\begin{abstract}
The problem of base station cooperation has recently been set within the framework of Stochastic Geometry. Existing works consider that a user dynamically chooses the set of stations that cooperate for his/her service. However, this assumption often does not hold. Cooperation groups could be predefined and static, with nodes connected by fixed infrastructure. To analyse such a potential network, in this work we propose a grouping method based on proximity. It is a variation of the so called Nearest Neighbour Model. We restrict ourselves to the simplest case where only singles and pairs of base stations are allowed to be formed. For this, two new point processes are defined from the dependent thinning of a Poisson Point Process, one for the singles and one for the pairs. Structural characteristics for the two are provided, including their density, Voronoi surface, nearest neighbour, empty space and J-function. We further make use of these results to analyse their interference fields and give explicit formulas to their expected value and their Laplace transform. The results constitute a novel toolbox towards the performance evaluation of networks with static cooperation. 
\end{abstract}

\begin{keywords}
Cooperation; Static groups; Poisson cellular network; Thinning; Interference \end{keywords}

\section{Introduction}

Cooperation between base stations (BSs) is receiving in recent years a lot of attention, due to its potential to improve coverage and spectral efficiency. It has shown considerable benefits especially for cell edge users that suffer from inter-cell interference. It is also expected to play a significant role due to the coming densification of networks with HetNets \cite{DhillonBest12}. The concept of cooperation in the downlink implies that two or more BSs exchange user state information and data to offer a stronger beneficial signal with reduced interference. The total benefit depends on the amount of information exchanged, but also very importantly on the number and positions of nodes that take part in the cooperation.

Recent studies have approached the problem of downlink cooperation with the theory of Point Processes \cite{PPbre1981} and Stochastic Geometry \cite{BacBlaVol1}, where the network topologies follow a certain probability distribution. Often, BS positions are modelled by a Poisson Point Process (PPP) with some fixed density over the entire plane. Specifically, Baccelli and Giovanidis \cite{AGFBTWC14} have studied cooperation between pairs of BSs. Evaluation for any number of cooperating stations has been done by Nigam et al in \cite{HaenggiCo14}, Tanbourgi et al in \cite{TanbCoopJ14} and B{\l}aszczyszyn and Keeler in \cite{BlaGn14}. In all these works, the common ground is the use of PPPs and the fact that the cooperation is driven by the user, who defines the set of stations for his/her service.

However, the last assumption is not very realistic since it overburdens the backhaul/control channel with intensive communication between BSs. Furthermore, it is not very clear which user is served by which station. A simpler and more pragmatic approach is to define a-priori static groups of BSs (that do not change over time). In each group, the BSs may reliably communicate with each other by reservation of control channel bandwidth or installation of optical fibres between them. 
If the criterion for grouping relates to geographic proximity, BSs in a small distance will coordinate fast and will share a planar area of common interest. The important question is how exactly should these groups be defined?

The idea is not new and suggestions have already appeared by Papadogiannis et al \cite{Papad08}, Giovanidis et al \cite{GiovaWCNC12}, Akoum and Heath \cite{AkoumHeathJournal13} and Pappas and Kountouris \cite{PapKou14}. Static clusters have also been considered for the uplink in \cite{AndrCLUST09}, \cite{VenkaPIMRCa07} and \cite{GiannCoop14}. In these works however, groups are formed neither systematically, nor optimally. Other works model the problem of dynamic clustering as a coalition game \cite{MoJoMISO14}.

We propose in our work a criterion for BS grouping that only depends on geometry. It is a variation of the Nearest Neighbour Model for point processes suggested by H\"aggstr\"om and Meester \cite{HagMeest96}. According to this, two BSs belong to the same group if one of the two is the nearest neighbour of the other. This assumption is reasonable for the telecommunication networks because it forms groups based on proximity. The variation we consider here limits the maximum number of elements $K$ that can group together. Since the general problem is very complicated, after presenting the criterion in Section \ref{SecII}, we focus on the case $K=2$, where the BSs can be either single or cooperate in pair with another BS. This is already interesting, as it raises all the important questions of the general problem. It leads to fundamental results and is very challenging to study.

In Section \ref{SecIII} we formally give the notions of single BSs and pairs and define the two point processes $\Phi^{(1)}$ and $\Phi^{(2)}$ that result from a dependent thinning of  the original process $\Phi$. The total interference of the network results from the sum of the interferences of the individual processes $\Phi^{(1)}$ and $\Phi^{(2)}$. In Section \ref{SecIV} we show that these processes are not PPPs and provide many of their structural properties: the average proportion of atoms from $\Phi$ that belong to $\Phi^{(1)}$ and $\Phi^{(2)}$, the average proportion of Voronoi surface related to each of them, as well as properties concerning repulsion/attraction. We continue in Section \ref{SecV} with the interference analysis and obtain explicit expressions for the expected value of the interference created by each one of the two point processes. We also provide their Laplace Transform (LT) when they are constrained within a finite subset of $A\subset\mathbb{R}^2$. Finally, Section \ref{SecV} concludes our work. All proofs of theorems can be found in the Appendix.

\section{Organising Base Stations into Groups}
\label{SecII}
Let us consider a Poisson Point Process (PPP) $\Phi$ in $\mathbb{R}^2$ with density $\lambda>0$.
A realisation $\phi$ of the process can be described by the infinite set of (enumerated) atoms $\left\{x_i\right\}$. Each realisation represents a possible deployment of single antenna Base Stations (BSs) on the plane. We wish to organise these BSs (or atoms) into \textit{cooperative groups} $\mathcal{C}_m\left(\phi\right)$, with possibly different sizes, where size refers to group cardinality $card\left(\mathcal{C}_m\right)$. The index $m$ enumerates the formed groups. We consider groups of atoms whose union exhausts the infinite set $\phi$ and they are \textit{disjoint}
\begin{eqnarray}
\label{ClusterProp1}
\bigcup_{m=1}^{\infty} \mathcal{C}_m & = & \phi,\\
\label{ClusterProp2}
\mathcal{C}_m\cap\mathcal{C}_n & = & \emptyset,\ \ \forall m\neq n.
\end{eqnarray}

We aim to find groups that are invariable in size and elements with respect to the random parameters of the telecommunication network (e.g. fading, shadowing or user positions). In this sense, we look for a criterion that aims at \textit{network-defined, static} clusters that differ from the user-driven selection of previous works.
For this reason, we use rules that depend only on geometry. Based on these, an atom $x\in\phi$  takes part in a group, based solely on its relative distance to the rest of the atoms $\phi\setminus\left\{x\right\}$. This geometric criterion is related to the path-loss factor of the channel gain. When a user lies at a planar point $z$ and is served by BS $x$, the gain is equal to
\begin{eqnarray}
\label{channelH}
h\left(z,x\right) = \nu\left(z,x\right)d\left(z,x\right)^{-\beta}, & & z\in\mathbb{R}^2,\ x\in\phi,
\end{eqnarray}
where $d(z,x):=\left|z-x\right|$ is the Euclidean distance between the user and the BS, $d\left(z,x\right)^{\beta}$ is the path loss with exponent $\beta>2$ and $\nu\left(z,x\right)$ models the power of channel fading. Both $h$ and $\nu$ hence refer to power, while the channel fading component is a complex number equal to $\sqrt{\nu}e^{j\theta}$ (e.g. for Rayleigh fading $\nu$ is exponentially distributed). 

\subsection{The Nearest Neighbour Model.}
In our work we investigate grouping decisions based on a model proposed by H\"aggstr\"om and Meester \cite{HagMeest96}, and further analysed in \cite{DaleyLast05}, \cite{DalStoSto99}, \cite{Kozakova06}, the so called \textit{Nearest Neighbour Model} (NNM). Given the realization $\phi$ we connect each atom $x$ to its geometrically \textit{Nearest Neighbour} by an undirected edge. This results in a graph $\mathcal{G}_{NN}$, which is well defined, because for a PPP no two inter-atom distances are the same a.s. and hence each atom has a unique first neighbour. However, an atom can be the nearest neighbour for a set of atoms (possibly empty).

\begin{figure}[t!]
	\centering  
	\includegraphics[trim = 40mm 75mm 40mm 70mm, clip, width=0.36\textwidth]{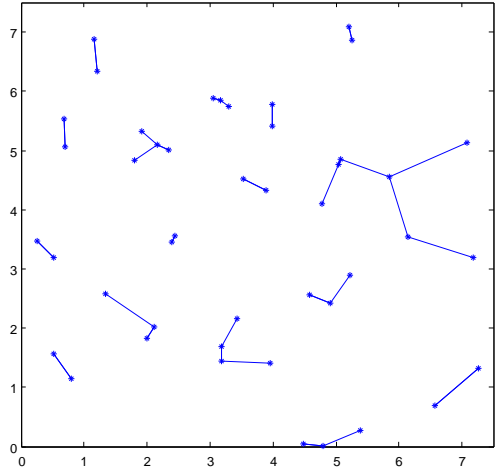}
       	\caption{Example of cooperation groups using the Nearest-Neighbour Model}
	\label{COOPclustNN}
\end{figure}

The NNM has certain properties that make it a good candidate for our purpose. (P.1) The group formation is independent of the PPP density $\lambda$. (P.2) The graph $\mathcal{G}_{NN}$ is disconnected, i.e. there always exist two atoms not connected by any path. (P.3) Each resulting cluster $\mathcal{C}$ does not contain \textit{cycles}, it is a \textit{tree} and hence the graph $\mathcal{G}_{NN}$ is a \textit{forest}. (P.4) The graph contains a.s. no infinite component, i.e. it does not percolate \cite[Th.2.1 and Th.5.2]{HagMeest96}. Consequently, the cardinality of each cluster is a.s. finite. (P.5) All atoms necessarily have a nearest neighbour. An example of a $\mathcal{G}_{NN}$ for a realisation of a PPP within a fixed window is shown in Fig.\ref{COOPclustNN}.

\textbf{Model Variation:} Although the NNM guarantees that the groups are finite, there is no upper bound on the group size. When referring to a telecommunications network, however, it is often more reasonable to bound the maximum group size by a number $K$. For $K=2$ this means that there can appear only single atoms and pairs, for $K=3$ singles, pairs and triplets etc. It is this modification of the NNM that we propose and analyse in this work, because it will lead to more natural grouping of BSs. Algorithmically, starting by a realisation of single atoms, we first search for possible groups of two (pairs). These are formed when two atoms are mutually nearest neighbours of each other and we result in the case $K=2$. From this last constellation, to shift to the case $K=3$, we will consider the existing pairs. For each pair we will search among the single atoms to find if there exist some of them that have one of the pair as nearest neighbour. If so, we choose the geographically closest with this property and create a group of three. Observe that not all of the pairs will become triplets. We iterate in this way for larger $K$.

\section{The special case of NNM groups with $K=2$.}
\label{SecIII}
From this point on, the paper will be devoted to the study of the simplest of static cooperation cases, the case $K=2$. 

\subsection{Singles and Pairs.}
\label{SinglePair}
The following definitions may apply to any process $\Phi$. In our work we will analyse the case where $\Phi$ is a PPP. For two different atoms $x,y\in\phi$, if $x$ is in \textit{Nearest Neighbour Relation (NNR)} with $y$, that is, if 
\begin{eqnarray*}
y=\argmin_{z\in \phi\setminus \{x\} } d(x,z),
\end{eqnarray*}  
we write $x \overset{\phi}{\rightarrow} y$. If this is not true we write  $x\overset{\phi}{\nrightarrow} y$. We use $\Phi$ instead of $\phi$, when we  consider the whole set of realisations $\Omega$ ($\omega\in\Omega$, $\Phi\left(\omega\right)=\phi$), i.e. when we calculate probabilities. We will omit the dependence on $\Phi$ or $\phi$ when it is clear from the context.
\begin{defi}
\label{defi1}
Two atoms $x,y\in \phi$, are in \textit{Mutually Nearest Neighbor Relation (MNNR)} if, and only if, $x \overset{\phi}{\rightarrow} y$ and $y \overset{\phi}{\rightarrow} x$. We denote this by $x \overset{\phi}{\leftrightarrow} y$. The two atoms then form a pair. In telecommunication terms we say that BSs $x$ and $y$ are in cooperation.
\end{defi}

\begin{defi}
\label{defi2}
An atom $x$ is called single if it is not in MNNR (does not cooperate) with any other atom in $\phi$. (Formally if for every $y\in \phi \setminus \{x\}$ such that $x\overset{\phi}{\rightarrow} y$, it holds that $y\overset{\phi}{\nrightarrow} x$). We denote this atom by $x _{\#}^{\phi}$. We then say that BS $x$ transmits individually.
\end{defi}

\begin{figure}[ht!]
	\centering
		\subfigure[The atoms $x$ and $y$ are in \textit{MNNR} $x \overset{\phi}{\leftrightarrow} y$ (i.e. $x \rightarrow y$, and $y\rightarrow x$).]{\includegraphics[trim = 20mm 50mm 5mm 50mm, clip, width=0.23\textwidth]{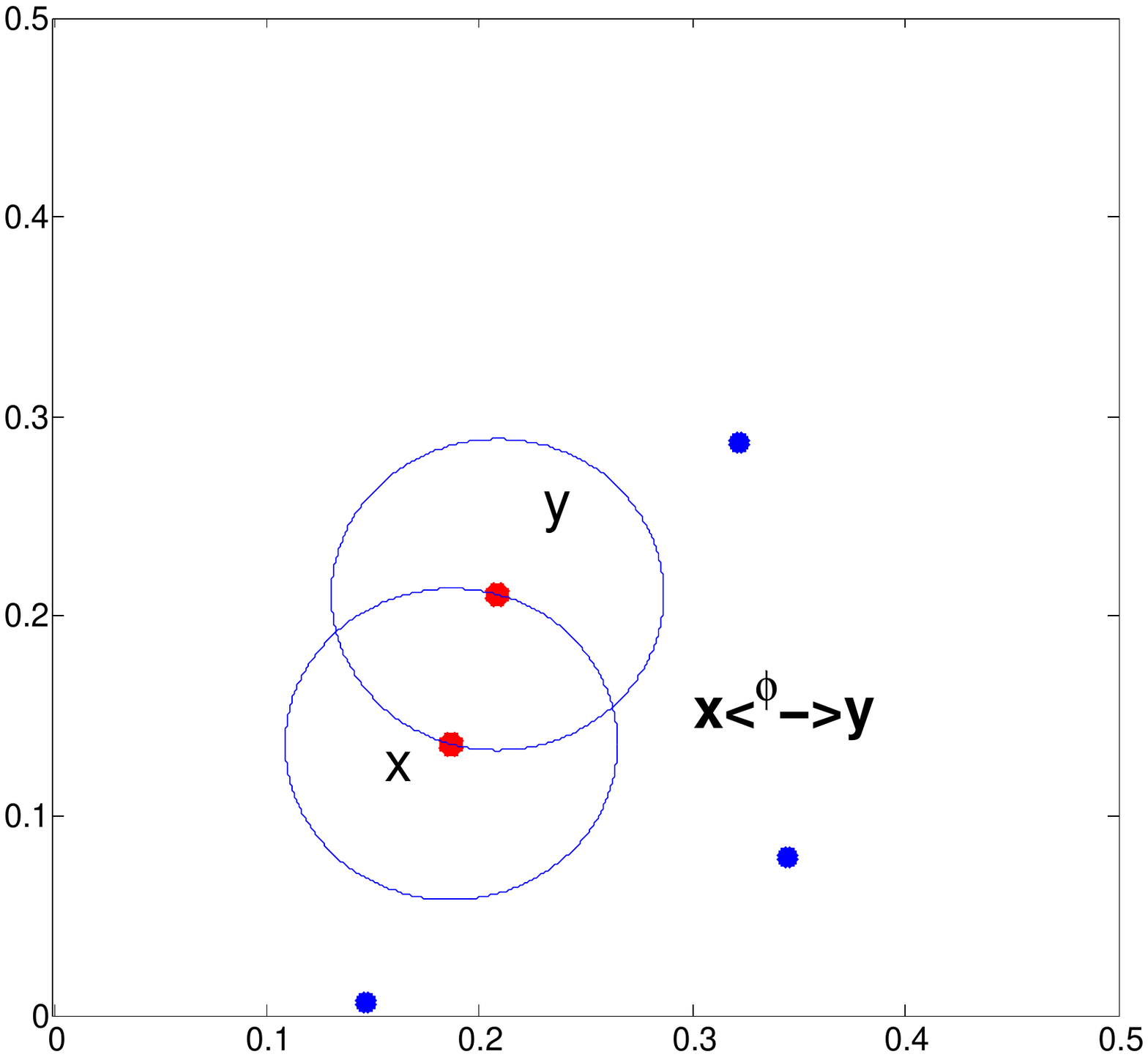}
	\label{SGD0:1}
	}
	\subfigure[The atom $w$ is a single and we write $w_{\#}^{\phi}$ (i.e. $w \rightarrow y$, but $y\nrightarrow w$).]{\includegraphics[trim = 20mm 50mm 5mm 50mm, clip, width=0.23\textwidth]{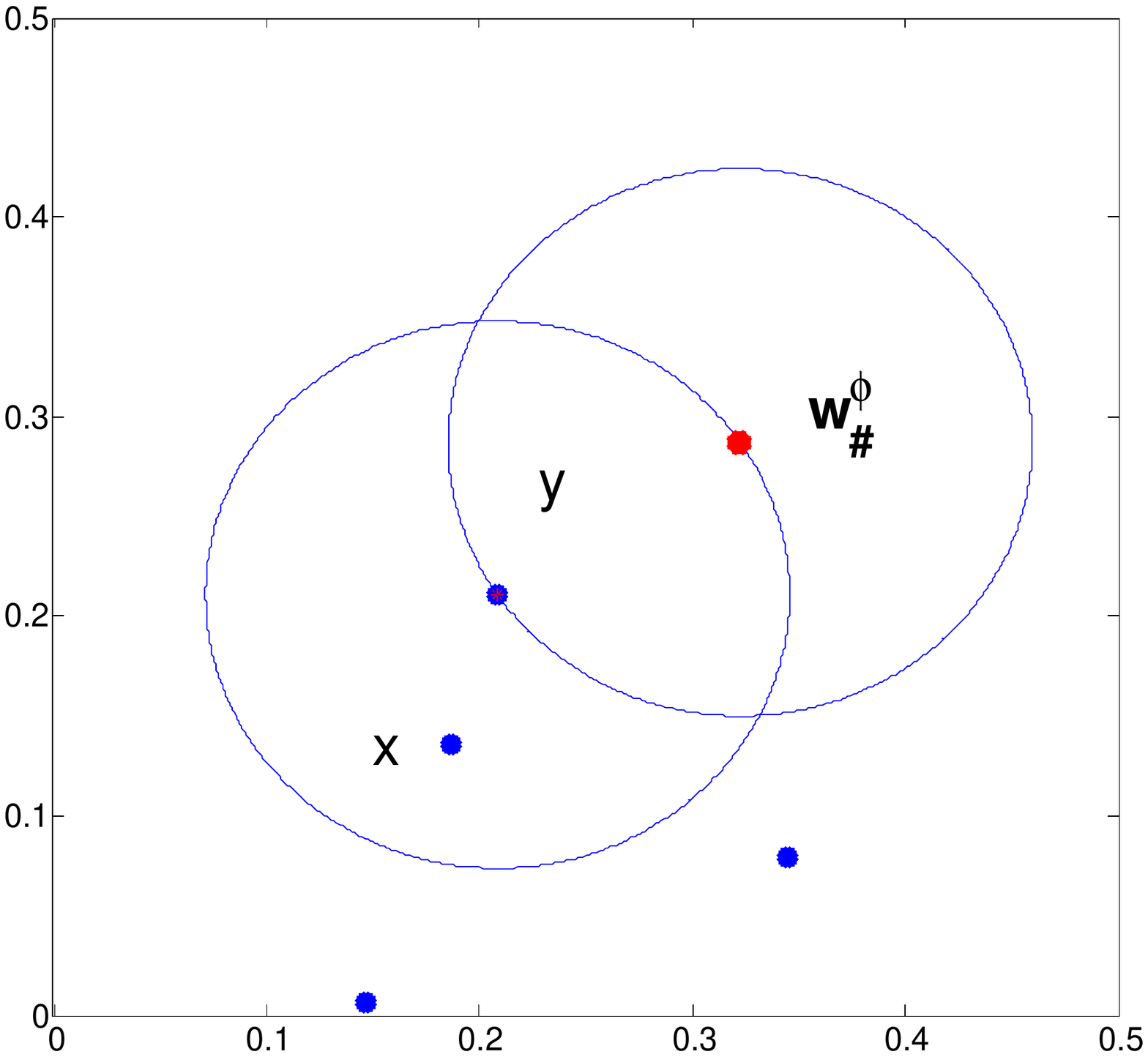}
	\label{SGD0:2}
	}
        \caption{Illustration of a pair of atoms and a single atom.}
\end{figure}

In geometric terms, the NNR $x \overset{\phi}{\rightarrow} y$ holds if and only if (iff) there exists a disc $\mathcal{B}\left(x,r\right)$ with radius $r=\left|x-y\right|$ empty of atoms in $\phi$. The nearest neighbour of $x$ is \textit{unique} in the case of PPP, because the probability of finding more than one atom on the circumference is zero. Furthermore, the MNNR holds, iff a symmetric relation is true i.e. the disc $\mathcal{B}\left(y,r\right)$ centred at $y$ with the same radius is empty. We conclude that the MNNR holds for $x,y$, iff the area $C\left(x,y\right):=\mathcal{B}\left(x,\left|x-y\right|\right)\cup\mathcal{B}\left(y,\left|x-y\right|\right)$ is empty of atoms. Its surface $\mathcal{S}$ is equal to $\mathcal{S}\left(C(x,y)\right)=\pi\left|x-y\right|^2\left(2-\gamma\right)$. Here, $\gamma:=\frac{2}{3}-\frac{\sqrt{3}}{2\pi}\approx 0.391$ \cite{DalStoSto99} is a constant number equal to the surface, divided by $\pi$, of the intersection of two discs with unit radius and centres lying on the circumference of each other. An illustration of the above explanations is given in figure \ref{SGD0:1}. On the other hand, figure \ref{SGD0:2} illustrates the single atom definition, where for any $y$ for which $x \overset{\phi}{\rightarrow} y$, the disc with centre $y$ and radius $\left|x-y\right|$ contains at least one atom of the process. 

With the above, and the empty space function for PPPs \cite{BacBlaVol1}, we find the probability of two atoms being in pair. 
\begin{lem} 
\label{Lemma1}
Given a PPP $\Phi$ with density $\lambda$ and conditioned that at $x,y\in\mathbb{R}^2$ there are two of its atoms, the probability that these are in MNNR is equal to
\begin{eqnarray}
\mathbb{P}\left(x \overset{\Phi}{\leftrightarrow} y\right) & = & e^{-\lambda\pi |x-y|^2(2-\gamma)},
\end{eqnarray}
where $\gamma :=  \frac{2}{3}-\frac{\sqrt{3}}{2\pi}$.
\end{lem}

Next, we give the global probability of any atom of $\Phi$ being single or in a pair. This result is shown from the point of view of an atom at $x$, i.e. involving the Palm measure $\mathbb{P}^{x}$ (see Appendix). We can  heuristically understand the Palm measure as $\mathbb{P}^{x}(\cdot)=\mathbb{P}(\cdot | x\in \Phi)$ \cite{BaddNotes07}. This description explains well its use, but the probability of $\left\{x\in\Phi\right\}$ is always null. Still we can assume that there exists a point of $\Phi$ within a small neighbourhood around $x$ (say a ball with radius $\epsilon>0$ and centre $x$) and then calculate the required probability, taking the limit $\epsilon\rightarrow 0$.
\begin{theo}
\label{Percentage}
Given a PPP $\Phi$ with density $\lambda$ and $x\in \Phi$, there exists a constant $p^*$, independent of $\lambda$ and $x$, such that 
\begin{equation}
\begin{split}
& \mathbb{P}\left( x \overset{\Phi}{\leftrightarrow} y, \mbox{ for some } y\in  \Phi \right)=p^*.\\
& \mathbb{P}\left( x_{\#}^{\Phi} \right) = 1-p^*.
\end{split}
\end{equation}
Specifically, $p^* = \frac{1}{2-\gamma}\approx 0.6215$.
\end{theo}
We should remark that Theorem \ref{Percentage} gives also the percentage of points that are singles or in pair, for any planar area $A\subseteq\mathbb{R}^2$. This is independent of the point of view of an atom at $x$. It means that, in average, a percentage $37.85\%$ of atoms of $A\cap\Phi$ are singles and $62.15\%$ of atoms are in pair. This has been verified by Monte Carlo simulations over a finite (but large enough) window $A$. It can also be analytically evaluated, using $f(x)=\mathbbm{1}_{\left\{x\in A\right\}}$ in the expression (\ref{EPhi1}), later on. The MNNR criterion leads, hence, to a reasonable splitting of the initial process into two processes, one with singles and another one with pairs of cooperating BSs.

Now, we can define two new point processes $\Phi^{(1)}$ and $\Phi^{(2)}$ that result from the \textit{dependent thinning} of the PPP $\Phi$ with the MNNR criterion, using Definitions \ref{defi1} and \ref{defi2}
\begin{equation*}
\begin{split}
& \Phi^{(1)}=\{x\in \Phi\  \&\  x\  \text{is single} \}, \\ 
& \Phi^{(2)}=\{x\in \Phi\  \&\  x \ \text{cooperates with another element of}\ \Phi\}.
\end{split}
\end{equation*}


\subsection{Voronoi Cells.}
\label{VoronoiCells}

It follows naturally to investigate the size of Voronoi cells associated with single atoms or pairs. A Voronoi cell of atom $x\in\phi$ is defined to be the geometric locus of all planar points $z\in\mathbb{R}^2$ closer to this atom than to any other atom of $\phi$ \cite{CompGeomBook}. In a wireless network the Voronoi cell is important when answering the question, which users should be associated with which station. Let $z\curvearrowright \phi^{(1)}$ (resp. $z\curvearrowright\phi^{(2)}$) denote the event that $z$ belongs to the Voronoi cell of some atom of $\phi^{(1)}$ (resp. $\phi^{(2)}$).  For the probability of these events we have analytical forms but not their numerical solutions, due to numerical issues related to integration over multiple overlapping circles.

\begin{figure}[t!]
	\centering  
	\includegraphics[trim = 15mm 65mm 10mm 65mm, clip, width=0.35\textwidth]{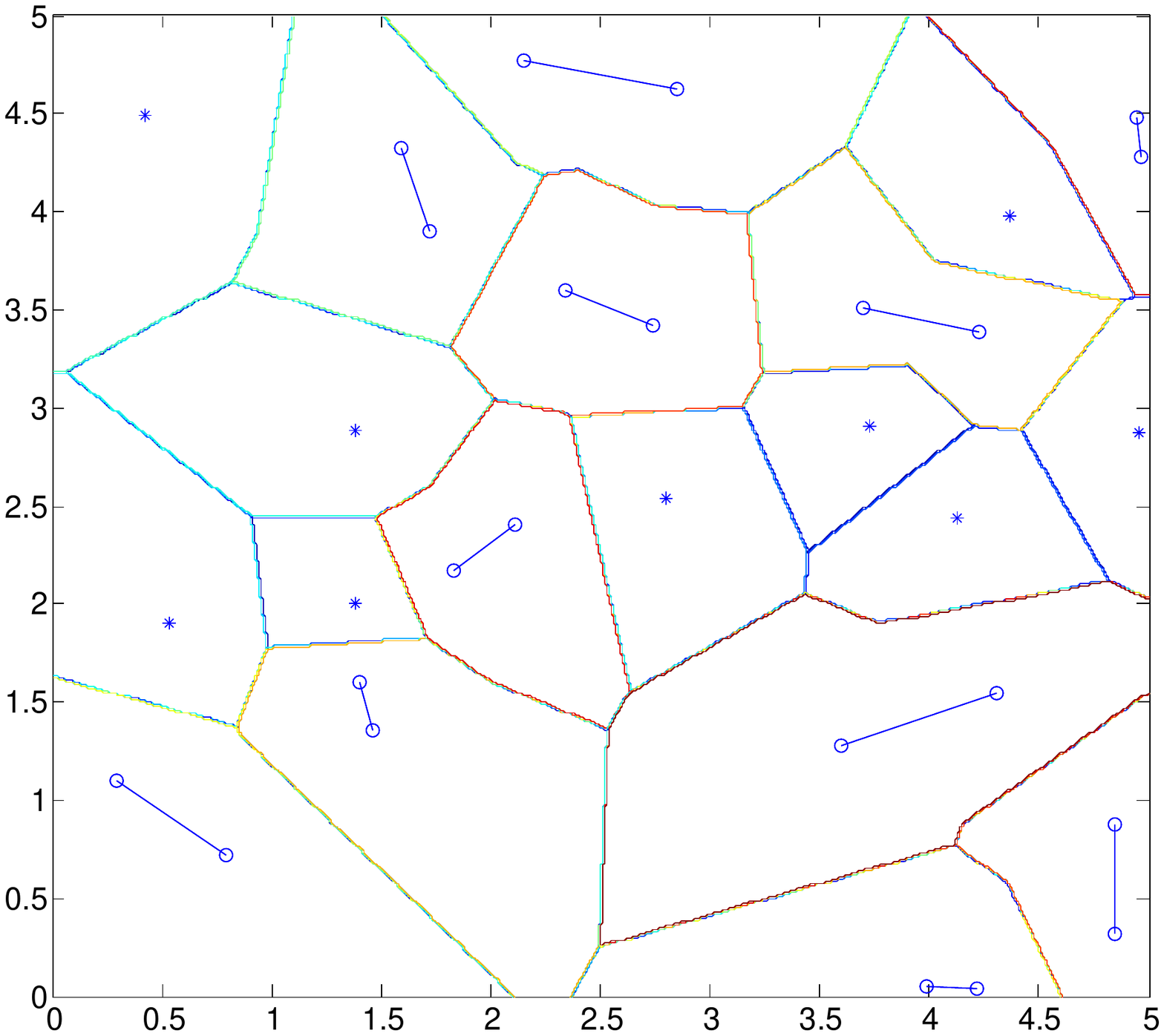}
       	\caption{Example of planar association areas for singles and pairs for a realisation of a PPP in a square window of size $t=5$. It is based on the Voronoi tessellation. 
	}
	\label{SGDP}
\end{figure}

\begin{NR}
\label{NR1}
The average surface proportion of Voronoi cells associated with single atoms and that associated with pairs of atoms is independent of the parameter $\lambda$. By Monte Carlo simulations, we find these values equal to
\begin{eqnarray}
\label{MCs}
\mathbb{P}\left(z\curvearrowright \Phi^{(1)}\right) & \approx & 0.4602.\\
\label{MCd}
\mathbb{P}\left(z\curvearrowright \Phi^{(2)}\right) & \approx & 0.5398.
\end{eqnarray}
\end{NR}
Interestingly, although the ratio of singles to pairs is $0.3785/0.6215$ the ratio of associated surface is $0.4602/0.5398$, implying that the typical Voronoi cell of a single atom is larger that that of an atom from a pair. The last remark gives first intuition that there is attraction between the cooperating atoms and a repulsion among the single atoms. We will further analyse these observations in the next section. We close this paragraph giving an example of the association cells for the singles and the pairs in Fig.\ref{SGDP}. For the pairs we consider the union of Voronoi cells of their individual atoms, as total association cell. 

\section{Characteristics of $\Phi^{(1)}$ and $\Phi^{(2)}$.}
\label{SecIV}

In this section we consider each one of the two newly defined processes $\Phi^{(1)}$ and $\Phi^{(2)}$ separately and analyse their behaviour. Specifically, we try to establish possible similarities related to PPPs and for each one discuss issues of repulsion and attraction between their atoms.

\subsection{Non-Poissonian behaviour.}
A first property is that both processes are \textit{homogeneous}. This is due to the homogeneity of $\Phi$ and because by definition, $\Phi^{(2)}$ depends only on the distance between elements of $\Phi$. Similarly for $\Phi^{(1)}$.
Since the two processes result from a dependent thinning of a PPP $\Phi$ they may not be PPPs. In fact we can show that they are not: Suppose that $\Phi^{(2)}$ is a homogeneous PPP. As shown in Theorem \ref{Percentage}, the percentage of its atoms in \textit{MNNR} with some other atom of $\Phi^{(2)}$ should be around $62.15\%$. However, by definition, $100\%$ of the elements of $\Phi^{(2)}$ are in \textit{MNNR} and we conclude that $\Phi^{(2)}$ is not a PPP. For $\Phi^{(1)}$ the argumentation is not as simple. Nevertheless, we can show this with Monte Carlo simulations on the average number of single points, which is far from the number $1-p^*=37.85\%$, and also using the Kolmogorov-Smirnov test \cite{Sheskin07}, which shows that the number of $\Phi^{(1)}$ atoms within a finite window is not Poisson distributed.

In what follows, we will use the PPP as reference process for comparison with the two new processes. For this, we consider two independent PPPs, $\hat{\Phi}^{(1)}$ and $\hat{\Phi}^{(2)}$, that result from independent thinning of the original PPP with probability $1-p^*$ and $p^*$ respectively. This is motivated from Theorem \ref{Percentage}. These PPPs will have density $\lambda_1=\lambda(1-p^*)$ and $\lambda_2=\lambda p^*$, which gives the same average number of points as the processes $\Phi^{(1)}$ and $\Phi^{(2)}$.

\subsection{Nearest Neighbour function.}
The \textit{Nearest Neighbour function (NN)}, denoted by \textit{G}, is the cumulative distribution function (cdf) of the distance from a typical atom of the process to the nearest other atom of the process \cite{BaddNotes07}. We first analyse $\Phi^{(2)}$ and for this we denote by $\mathbb{P}^{(2),x}$, and $G^{(2)}(r)$ the associated Palm probability measure and \textit{NN} function of the distance $r\geq 0$, respectively. 

\begin{figure*}
\begin{tabular}{c c c}
		\subfigure[\textit{NN} of $\Phi^{(1)}$ (blue point) and $\hat{\Phi}^{(1)}$ (red dash).]{\includegraphics[trim = 15mm 65mm 10mm 65mm, clip, width=0.32\textwidth]{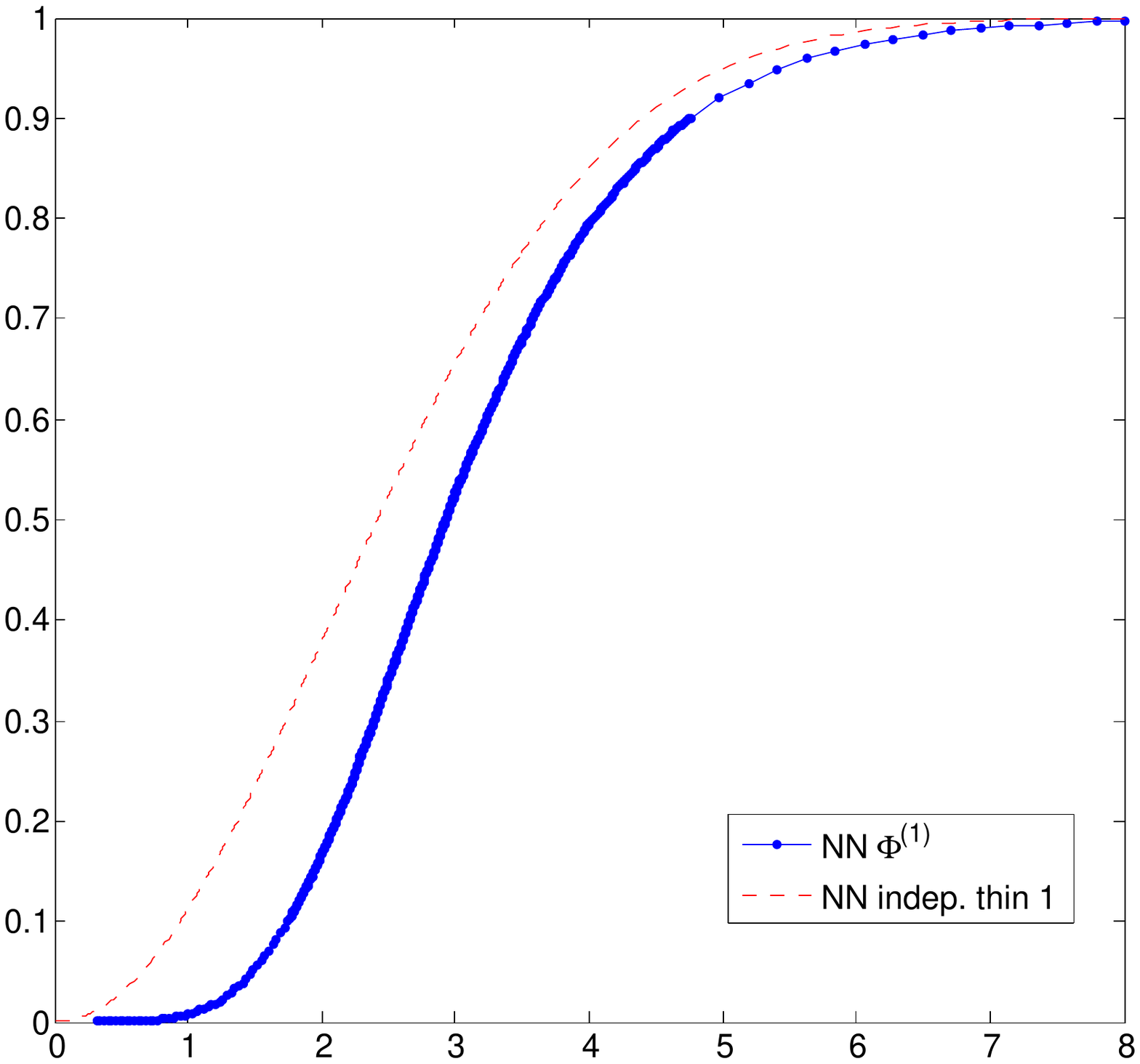}
		\label{fig:SGD2_2}
		}
		&
		\subfigure[\textit{ES} of $\Phi^{(1)}$ (blue point) and $\hat{\Phi}^{(1)}$ (red dash).]{\includegraphics[trim = 15mm 65mm 10mm 65mm, clip, width=0.32\textwidth]{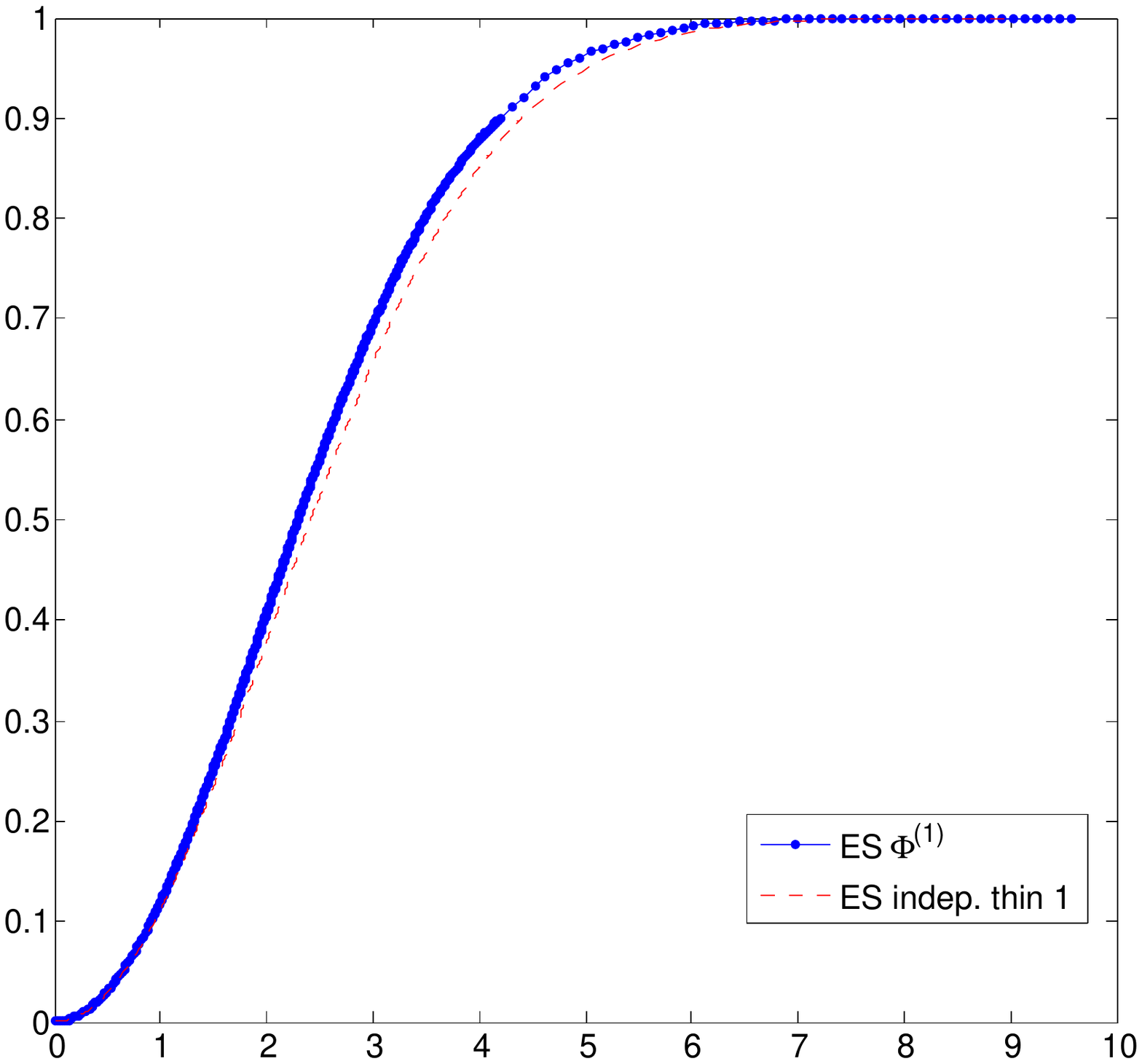}
		\label{fig:SGD3_1}
		}
		&
		\subfigure[\textit{J} function of $\Phi^{(1)}$.]{\includegraphics[trim = 18mm 57mm 10mm 55mm, clip, width=0.29\textwidth]{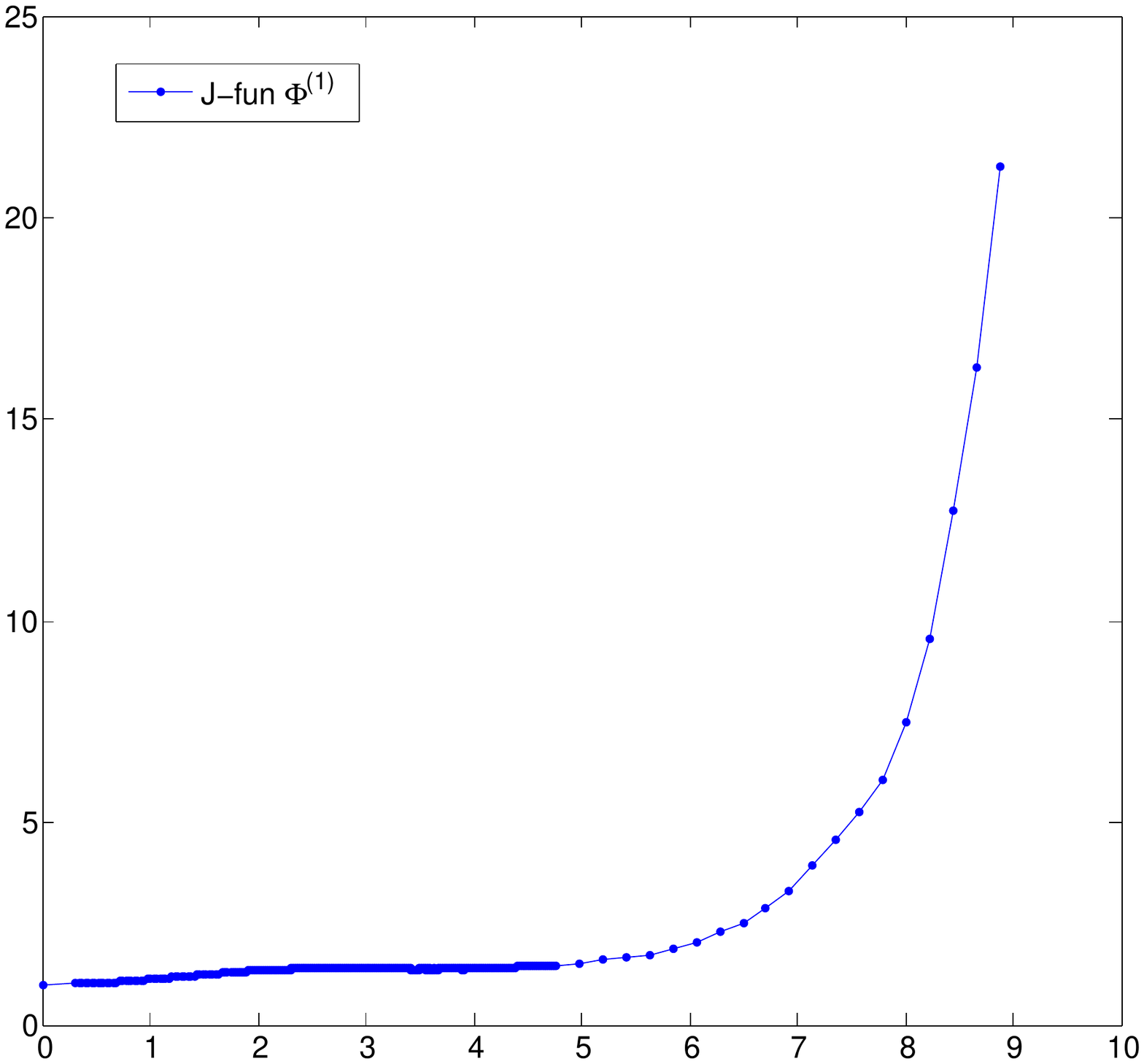}
		\label{fig:SGD5_1}
		}\\
		\subfigure[\textit{NN} of $\Phi^{(2)}$ (blue point) and $\hat{\Phi}^{(2)}$ (red dash).]{\includegraphics[trim = 15mm 65mm 10mm 65mm, clip, width=0.32\textwidth]{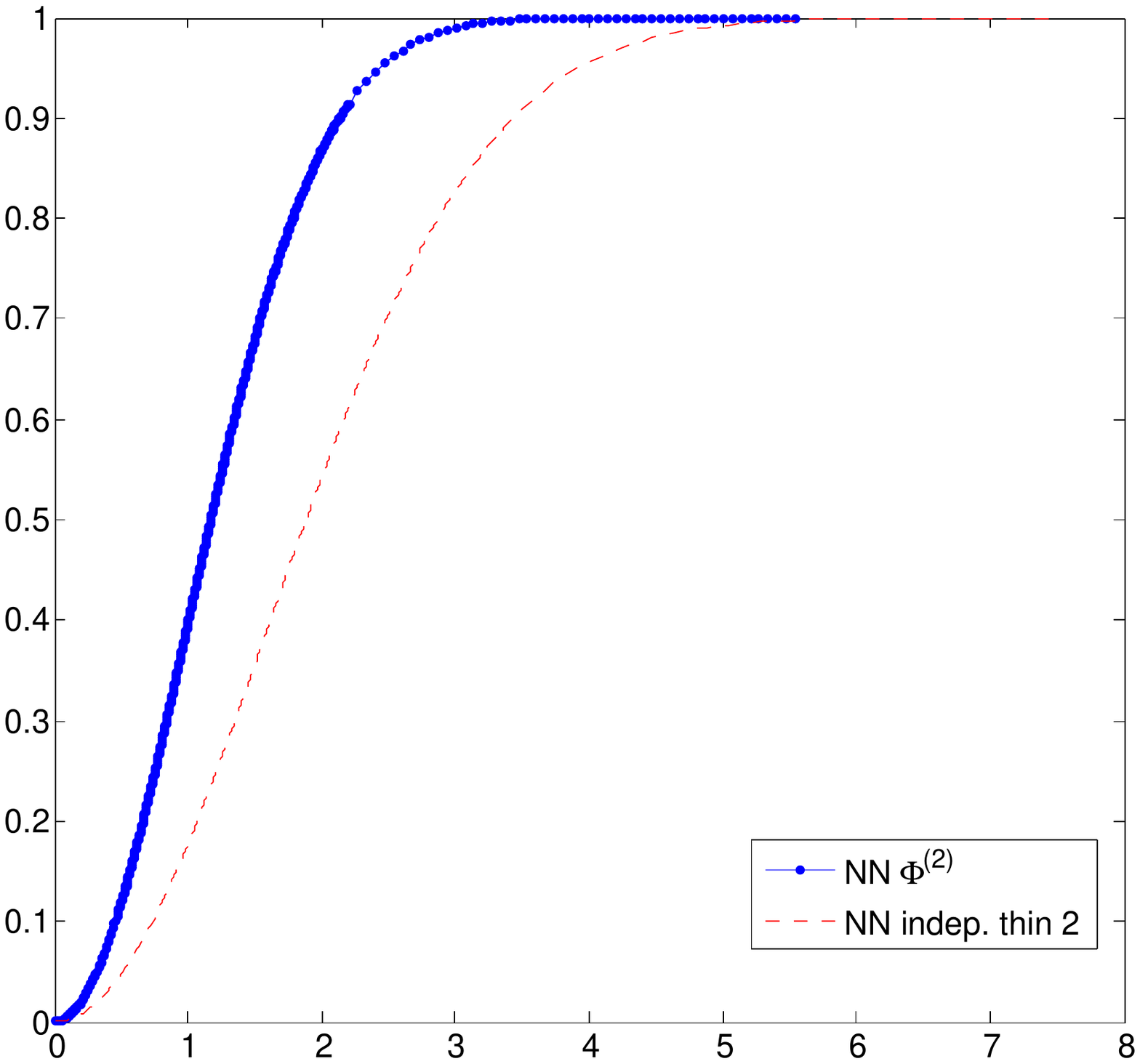}
		\label{fig:SGD2_1}
		}
		&
		\subfigure[\textit{ES} of $\Phi^{(2)}$ (blue point) and $\hat{\Phi}^{(2)}$ (red dash).]{\includegraphics[trim = 15mm 65mm 10mm 65mm, clip, width=0.32\textwidth]{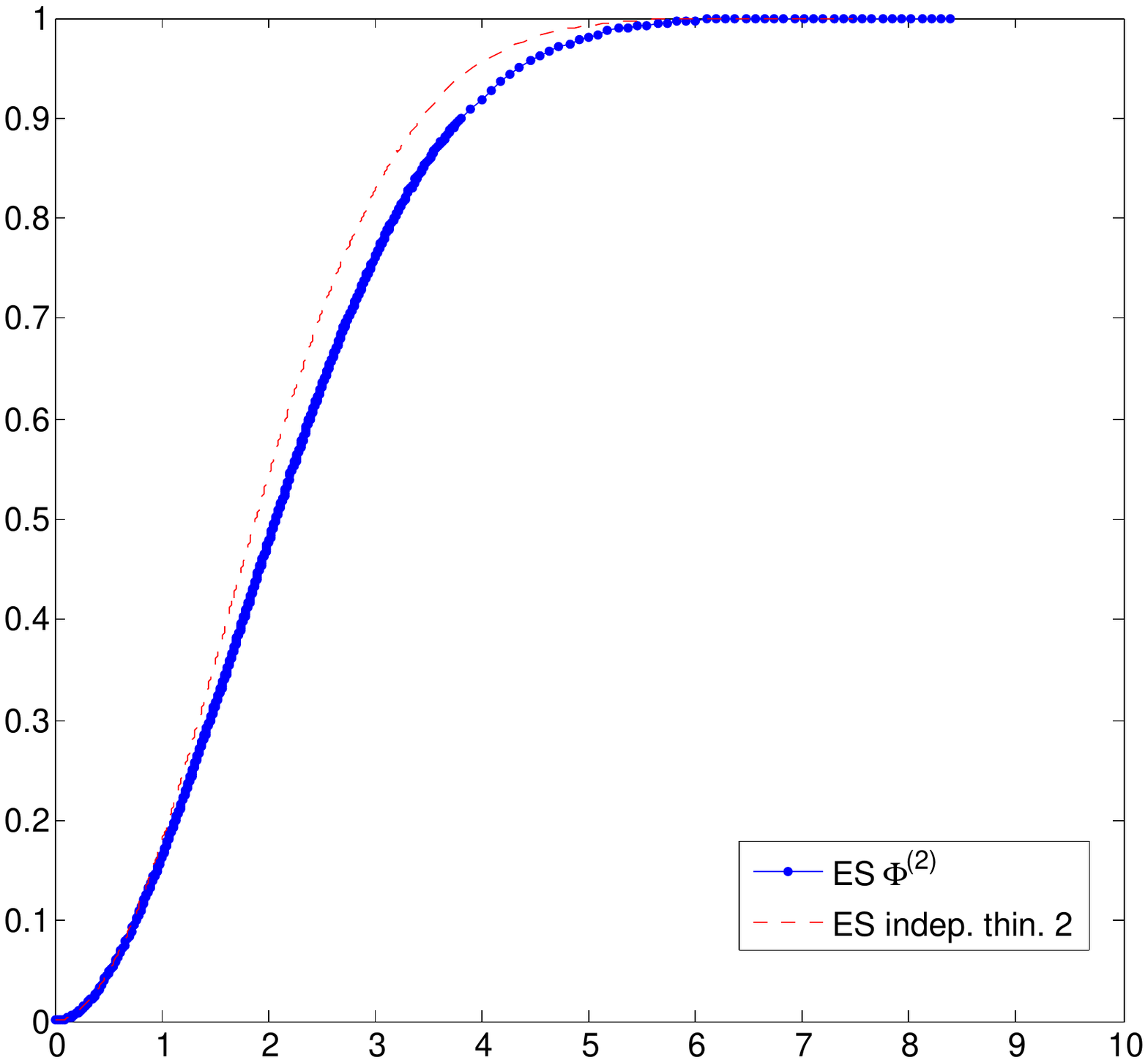}
		\label{fig:SGD3_2}
		}
		&
		\subfigure[\textit{J} function of $\Phi^{(2)}$.]{\includegraphics[trim = 18mm 65mm 10mm 65mm, clip, width=0.32\textwidth]{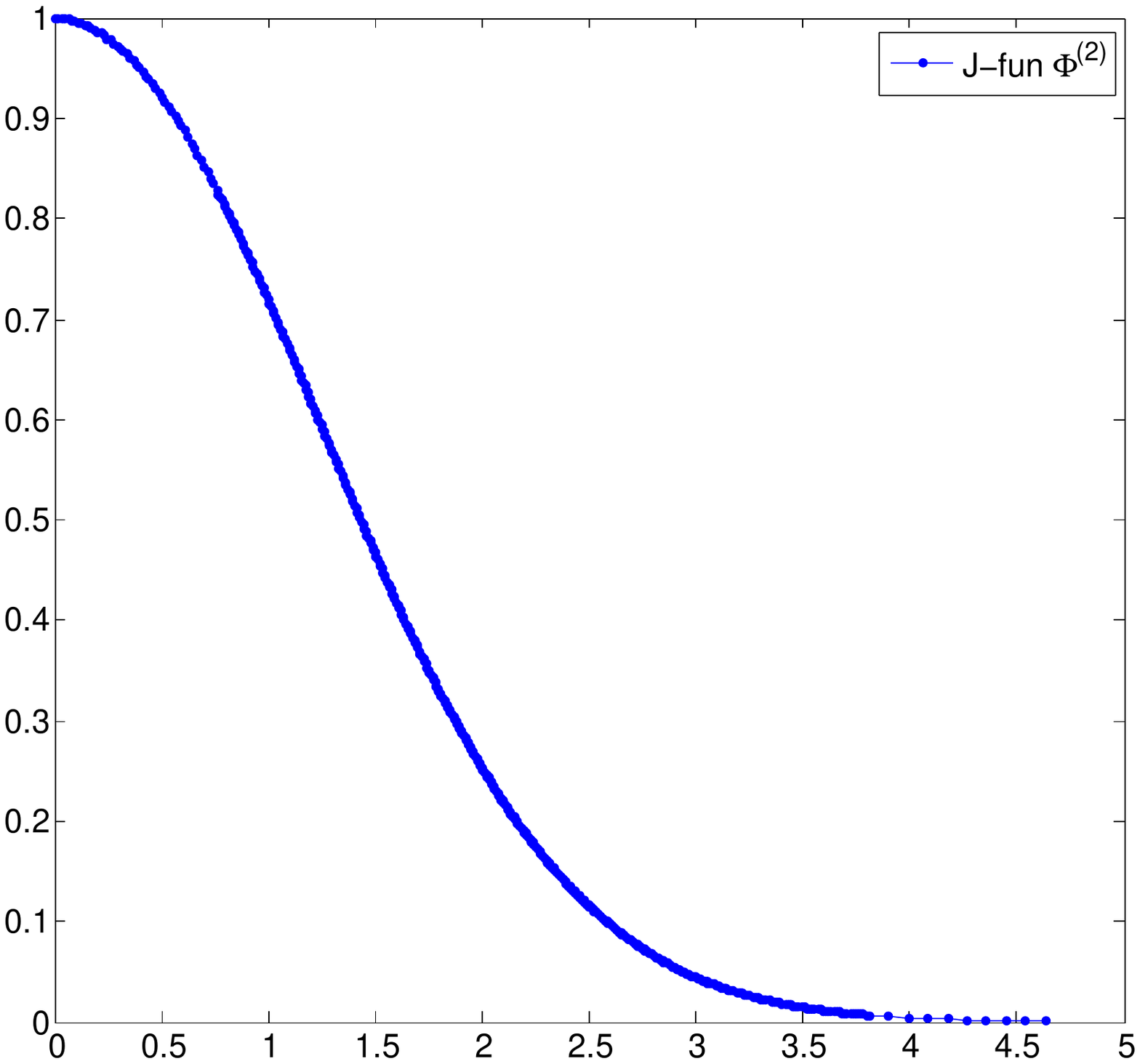}
		\label{fig:SGD5_2}
		}
	         \end{tabular}
	       	\caption{NN function, ES function and J function for the processes $\Phi^{(1)}$, $\Phi^{(2)}$ and their comparison with the PPPs $\hat{\Phi}^{(1)}$ and $\hat{\Phi}^{(2)}$. }	
\end{figure*}

\begin{theo}
\label{TheoNN2}
The NN function of $\Phi^{(2)}$ is equal to
\begin{eqnarray}
\label{NNPhi2}
G^{(2)}(r) & = & \mathbb{P}^{(2),x}(d(x,\Phi^{(2)}\setminus \{x\})\leq r)\nonumber\\
& = & 1-e^{-\lambda\pi r^2 (2-\gamma)},
\end{eqnarray}   
where $\gamma$ is the same constant as in Lemma \ref{Lemma1}.
\end{theo}
Hence, the NN random variable (r.v.) is Rayleigh distributed, with scale parameter $\sigma=(2\lambda\pi(2-\gamma)).^{-1/2}$. Unfortunately, we have not been able to provide a closed analytic expression for the \textit{NN} function of the process $\Phi^{(1)}$, as we did for $\Phi^{(2)}$ in equation \eqref{NNPhi2}. In Figures \ref{fig:SGD2_2} and \ref{fig:SGD2_1} we plot the NN function of $\Phi^{(1)}$ using Monte Carlo simulations and $\Phi^{(2)}$ from the analytic expression in (\ref{NNPhi2}), respectively. Both plots are compared to the NN functions of the independent PPPs $\hat{\Phi}^{(1)}$ and $\hat{\Phi}^{(2)}$. These are equal to (for $i\in\left\{1,2\right\}$)
\begin{eqnarray}
\label{NNindiPhii}
\hat{G}^{(i)}(r) & = & \mathbb{P}^{x}(d(x,\hat{\Phi}^{(i)}\setminus \{x\})\leq r)\nonumber\\
& = & 1-e^{-\lambda_i\pi r^2}.
\end{eqnarray}   
To get an intuition how different the results are, observe that the exponent in the PPP case $\hat{G}^{(2)}$ depends on the density $\lambda_2=\lambda/(2-\gamma)$, whereas the expression for $G^{(2)}$ is the same but with exponent $\lambda\left(2-\gamma\right)$. This observation explains why the use of independent thinning would be an inappropriate approximation in our case. 

\subsection{Empty Space function.}

The \textit{empty space function (ES)}, denoted by \textit{F}, is the cdf of the distance from a fixed planar point $z\in \mathbb{R}^{2}$ to the nearest atom of the point process considered \cite{BaddNotes07}. By stationarity of $\Phi^{(1)}$ and $\Phi^{(2)}$, the function F does not depend on $z$, so we can consider the typical user (point) at the Cartesian origin. Unfortunately, we have not been able to derive analytical formulas for either of the two $F^{(1)}$ and $F^{(2)}$. The ES functions of the independent PPPs are denoted by $\hat{F}^{(1)}$ and $\hat{F}^{(2)}$ and they are equal to the expression of the NN functions in (\ref{NNindiPhii}). This is a property of the PPPs \cite{BaddNotes07}.

We have used Monte Carlo simulations to plot the ES of the two processes. In Fig.\ref{fig:SGD3_1} we show the comparison between $F^{(1)}$ and $\hat{F}^{(1)}$, which - unexpectedly - seem very close to each other. The same comparison in Fig.\ref{fig:SGD3_2} shows also closeness of fit, although less tight, between $F^{(2)}$ and $\hat{F}^{(2)}$. 
Since we have no analytic expressions, we are tempted to consider the ES function of the independently thinned PPPs as a reasonable approximation for those of $\Phi^{(1)}$ and $\Phi^{(2)}$.

\subsection{The \textit{J} function.}\label{Jfunction}

The two functions NN and ES can be combined into a single expression known as $J$ function. The latter is a tool introduced by van Lieshout and Baddeley \cite{BaddNotes07} to measure repulsion and/or attraction between the atoms of a point process. It is defined as
\begin{equation}
J(r)=\frac{1-G(r)}{1-F(r)}.
\end{equation} 
In the case of the uniform PPP, $G\left(r\right)\equiv F\left(r\right)$ and $J\left(r\right)=1$, as a consequence of the fact that the reduced Campbell measure is identical to the original measure. Hence the $J$ function quantifies the differences of any process with the PPP. When $J(r)>1$, this is an indicator of repulsion between atoms, whereas $J(r)<1$ indicates attraction. We plot in Fig.\ref{fig:SGD5_1} the \textit{J} function of $\Phi^{(1)}$ and in Fig.\ref{fig:SGD5_2} that of $\Phi^{(2)}$. From the figures we conclude \textit{that $\Phi^{(1)}$ exhibits repulsion for every $r\geq 0$, and $\Phi^{(2)}$ attraction everywhere}. However, note that the attraction in the case $\Phi^{(2)}$ is due to the pairs formed. If we consider a new process having as elements the pairs of $\Phi^{(2)}$, this process of pairs exhibits repulsion everywhere.
 

\section{Interference Analysis}
\label{SecV}

The purpose of the previous analysis was to develop the tools necessary for use in a communications context. Within this context, the cooperating BSs will have a different influence on the interference seen by a user in the network, than those operating individually. The current section will focus on the \textit{interference field} generated by $\Phi^{(1)}$ and $\Phi^{(2)}$. As shown in Section \ref{SecIV}, the two processes have a different behaviour compared to a PPP and this is why approximations based on independent thinning of $\Phi$ will not bring accurate results. We thus have to resort to a more direct approach.

We denote by $\mathcal{I}^{(1)}$ and $\mathcal{I}^{(2)}$, the interference field generated by $\Phi^{(1)}$ and $\Phi^{(2)}$, respectively. The typical user is  chosen at the Cartesian origin due to stationarity. To keep our setting as general as possible we describe by use of measurable functions $f:\mathbb{R}^2\longrightarrow \mathbb{R}^+$, $g:\mathbb{R}^2\times\mathbb{R}^2\longrightarrow \mathbb{R}^+$ the signals transmitted by a single BS or by a pair and received at the typical user. The sum over the entire (of each) process gives the random variables of interest
\begin{eqnarray}
\label{I1}
\mathcal{I}^{(1)} & = & \sum_{x\in \phi^{(1)}} f(x),\\
\label{I2}
\mathcal{I}^{(2)} & = & \frac{1}{2}\sum_{x\in \phi^{(2)}} \sum_{y\in \phi^{(2)}}^{\neq}g(x,y) \mathbbm{1}_{\left\{x \overset{\phi^{(2)}}{\leftrightarrow} y\right\}}.
\end{eqnarray} 
The $1/2$ in front of the summation in (\ref{I2}) prevents us from considering a pair twice, whereas $\neq$ implies that $y\in\phi^{(2)}\setminus\left\{x\right\}$. 

We now give some practical examples for the functions under study. First of all, the function for the individual BS in this analysis will be equal to
\begin{equation}
\label{fx}
f(x) = h(0,x)
\end{equation}
where $h(z,x)$ was defined in (\ref{channelH}). This is the signal from a single antenna BS that lies at distance $\left|0-x\right|$ from the user $0$. The fading power $\nu(0,x)$ follows the $\exp(1/P)$ distribution, where $P$ is the BS transmission power \cite{AndrewsCoverage}. The cooperation signal is more interesting. We can consider the following cases
\begin{eqnarray}
\label{gFUN}
g\left(x,y\right) = \left\{\begin{tabular}{l l}
$h(0,x)+h(0,y)$ & [NC]\\
$\max\left\{h(0,x),h(0,y)\right\}$ & [OF1]\\
$\mathbbm{1}_{on}h(0,x)+(1-\mathbbm{1}_{on})h(0,y)$ & [OF2]\\
$\left|\sqrt{h(0,x)}e^{i\theta_x}+\sqrt{h(0,y)}e^{i\theta_y}\right|^2$ &  [PH]
\end{tabular}\right..
\end{eqnarray}
[NC] refers to the no cooperation case, where both BSs of the pair behave individually. [OF1] refers to the case where the BS with the strongest interfering signal is actively serving its user, while the other is off (alternatively we could replace $\max$ by $\min$ to consider the weakest signal of the two). [OF2] is again a scenario with one of the two BSs active and the other off, where the choice is made randomly by a r.v. with $\mathbb{E}\left[\mathbbm{1}_{on}\right]=q$ (e.g. $q=0.5$ for fairness). Finally, [PH] is the case where the two complex signals are combined in phase as well \cite{AGFBTWC14}. Here, $i$ is the complex unit and $\theta_x,\theta_y$ are the uniformly random phases of the signals. Observe that the above signals can be generalised to include MIMO transmission as well.

\subsection{Expected value of $\mathcal{I}^{(1)}$ and $\mathcal{I}^{(2)}$.}

The next Theorem gives an exact integral expression to the expected value of the interference field generated by the singles and the pairs. The proof uses the Campbell-Little-Mecke formula, Lemma \ref{Lemma1}, and Theorem \ref{Percentage}.
\begin{theo}\label{Expected}
The expected value of the \textit{interference field} generated by $\Phi^{(1)}$ and $\Phi^{(2)}$ is given by 
\begin{align}
\label{EPhi1}
\mathbb{E} \left[\mathcal{I}^{(1)}\right] &= (1-p^*) \int_{\mathbb{R}^2}\mathbb{E}\left[f(x)\right] \lambda dx, \\
\label{EPhi2}
\mathbb{E} \left[\mathcal{I}^{(2)}\right] &= \frac{1}{2} \int_{\mathbb{R}^2} \int_{\mathbb{R}^2}\mathbb{E}\left[g(x,y)\right]e^{-\lambda\pi |x-y|^2(2-\gamma)}\lambda dy \lambda dx.
\end{align}
\end{theo}
The expected value can be finite or infinite, depending on the choice of $f(x)$ and $g(x,y)$. Observe that for [NC] and [PH] the expected interference has the same value.

\begin{cor}[Intensity measure]
\label{CorIM}
The intensity measures for $\Phi^{(1)}$ and $\Phi^{(2)}$, denoted by $M^{(1)}$ and $M^{(2)}$ respectively, are equal to
\begin{eqnarray}
M^{(1)}(dx) & = & (1-p^*)\lambda dx, \\
M^{(2)}(dx)& = & p^*\lambda dx.
\end{eqnarray}
\end{cor}


\subsection{Laplace transform of $\mathcal{I}^{(1)}$ and $\mathcal{I}^{(2)}$.}

As a final result we present our findings related to the Laplace transform (LT) of the interference from $\Phi^{(1)}$ and $\Phi^{(2)}$. We derive here the exact LT of the interference when considering a finite subset $A\subset\mathbb{R}^2$ (window) and the related random point measure
\begin{equation*}
\begin{split}
& \Phi^{(1)}_A  =\left\{
\mbox{
single atoms of $\Phi(A)=\Phi\cap A$ inside $A$.
}
\right\}\\
& \Phi^{(2)}_A =\left\{
\mbox{
atoms of $\Phi(A)$ in \textit{MNNR} with another atom in $A$.
}
\right\}\\
\end{split}
\end{equation*}
A sketch of proof is given due to space constraints. For finite subsets, the atoms of the PPP $\Phi$ are distributed i.i.d. uniformly within $A$. We condition on the number of atoms that appear within $A$, and make use of the fact that $\Phi$ is a PPP (or some other process with known counting measure). We also consider the MNNR only among the atoms in $A$ and not on the entire plane. Then by direct application of the expected value of a function $\chi$, we can write the LT as an infinite sum of terms $\chi(n)\mathbb{P}\left(N(A)=n\right)$, where $\chi(n)$ is the value of the function when $n$ atoms appear in $A$ and $\mathbb{P}\left(N(A)=n\right)$ is the probability that this event occurs. 

The method can be seen as an approximation of the LT of $\Phi^{(1)}$ and $\Phi^{(2)}$. It is an open question to prove that when $A\rightarrow\mathbb{R}^2$ the LT of the finite process converges to the LT we are looking for. We write $\Phi^{(i)}\approx \Phi^{(i)}_A$, for $A$ regular and large enough. However, since we are often interested in simulating and analysing only finite areas and also since the conjecture in the limiting case sounds reasonable, the result we present here has a great importance. It fully characterises the distribution of the interference (and is provable for finite windows).

For a finite number $n$ of known planar points $x_1,\ldots,x_n\in A$ (that are potentially occupied by atoms, when the latter are chosen uniformly within the area), we define the functions $H^{(n)}:\left(\mathbb{R}^2\right)^n\rightarrow (\mathbb{R}^+)^n$ and $J^{(n)}:\left(\mathbb{R}^2\right)^n\rightarrow (\mathbb{R}^+)^{n^2}$
\begin{equation*}
H^{(n)}_i(x_1,\ldots,x_n)=\begin{cases}
1, & \mbox{if} \ x_i \ \mbox{satisfies the "single" relation.}\\
0, & \mbox{otherwise.} 
\end{cases}
\end{equation*}
The "single" relation is satisfied in the deterministic case, exactly as in Definition \ref{defi2} if, for every $x_j\neq x_i$ for which $x_i \rightarrow x_j$ it holds that $x_j \nrightarrow x_i$. Then $H^{(n)}=(H_1^{(n)},\ldots,H_n^{(n)})^T$.
\begin{equation*}
J^{(n)}_{i}(x_1,\ldots,x_n)=\begin{cases}
(0\ldots 1\ldots), & \mbox{if} \ x_i \ \mbox{is in "pair" with}\ x_j.\\
(0\ldots0), & \mbox{ in other case.} 
\end{cases}
\end{equation*}
The "pair" relation is satisfied in the deterministic case, as in Definition \ref{defi1} if $x_i \leftrightarrow x_j$, for some $i\neq j$. Then $J^{(n)}=(J_1^{(n)},\ldots,J_n^{(n)})^T$, where $J^T$ is the transpose vector of $J$.

\begin{theo}[Laplace transform]
\label{LaplaceTransform}
Consider a PPP $\Phi$ with density $\lambda$, a subset $A\subset{\mathbb{R}^2}$ and the functions $f,g$ (e.g. (\ref{fx}) and (\ref{gFUN})) related to the single atoms and the pairs respectively. Let $F^{(n)}(x_1,\ldots,x_n)=(f(x_1),\ldots,f(x_n))$, $G_i^{(n)}(x_1,\ldots,x_n)=(g(x_i,x_1)\ldots g(x_i,x_n))$, and $G^{(n)}=(G_1^{(n)}\ldots G_n^{(n)})$.

The LT of the interference $\mathcal{I}^{(1)}$ for $\Phi^{(1)}_A$ is equal to 
\begin{equation*}
\begin{split}
& \mathbb{E} \left[ e^{-s\mathcal{I}^{(1)}} \right] = e^{-\lambda\mathcal{S}(A)} \Bigg(1 + \lambda\int_A \mathbb{E}\left[e^{-sf(x)}\right] dx + \frac{\lambda^2}{2}+ \\
& +\sum^ \infty_{n=3} \frac{\lambda^n}{n!}\int_A \ldots \int_A \mathbb{E}\left[e^{-sF^{(n)}\cdot H^{(n)}}\right] dx_1 \ldots dx_n \Bigg),
\end{split}
\end{equation*}

The LT of the interference $\mathcal{I}^{(2)}$ for $\Phi^{(2)}_A$ is equal to
\begin{equation*}
\begin{split}
& \mathbb{E}\left[e^{-s\mathcal{I}^{(2)}}\right]= e^{-\lambda \mathcal{S}(A)} \Bigg(1 + \lambda \mathcal{S}(A) \\
& +\frac{\lambda^2}{2} \int_A\int_A \mathbb{E}\left[e^{-\frac{s}{2}(g(x,y)+g(y,x))}\right]\lambda dy \lambda dx \\
& +\sum^ \infty_{n=3} \frac{\lambda^n}{n!}\int_A \ldots \int_A \mathbb{E}\left[e^{-\frac{s}{2}G^{(n)}\cdot J^{(n)}}\right]dx_1 \ldots dx_n \Bigg).
\end{split}
\end{equation*}
\end{theo}

We finish this section by evaluating the expected value of the interference using Theorem \ref{Expected} and the proposed expressions for $f(x)$ in (\ref{fx}) and for $g(x,y)$ in (\ref{gFUN}). We further compare the results from the numerical integration with those by Monte Carlo simulations within a finite - but large enough - window. The chosen density is $\lambda=0.1$ [atoms/$m^2$] and the window is a square of size $100\times 100$ [$m^2$].

\begin{figure}[t!]
	\centering  
	\includegraphics[trim = 0mm 50mm 5mm 60mm, clip, width=0.35\textwidth]{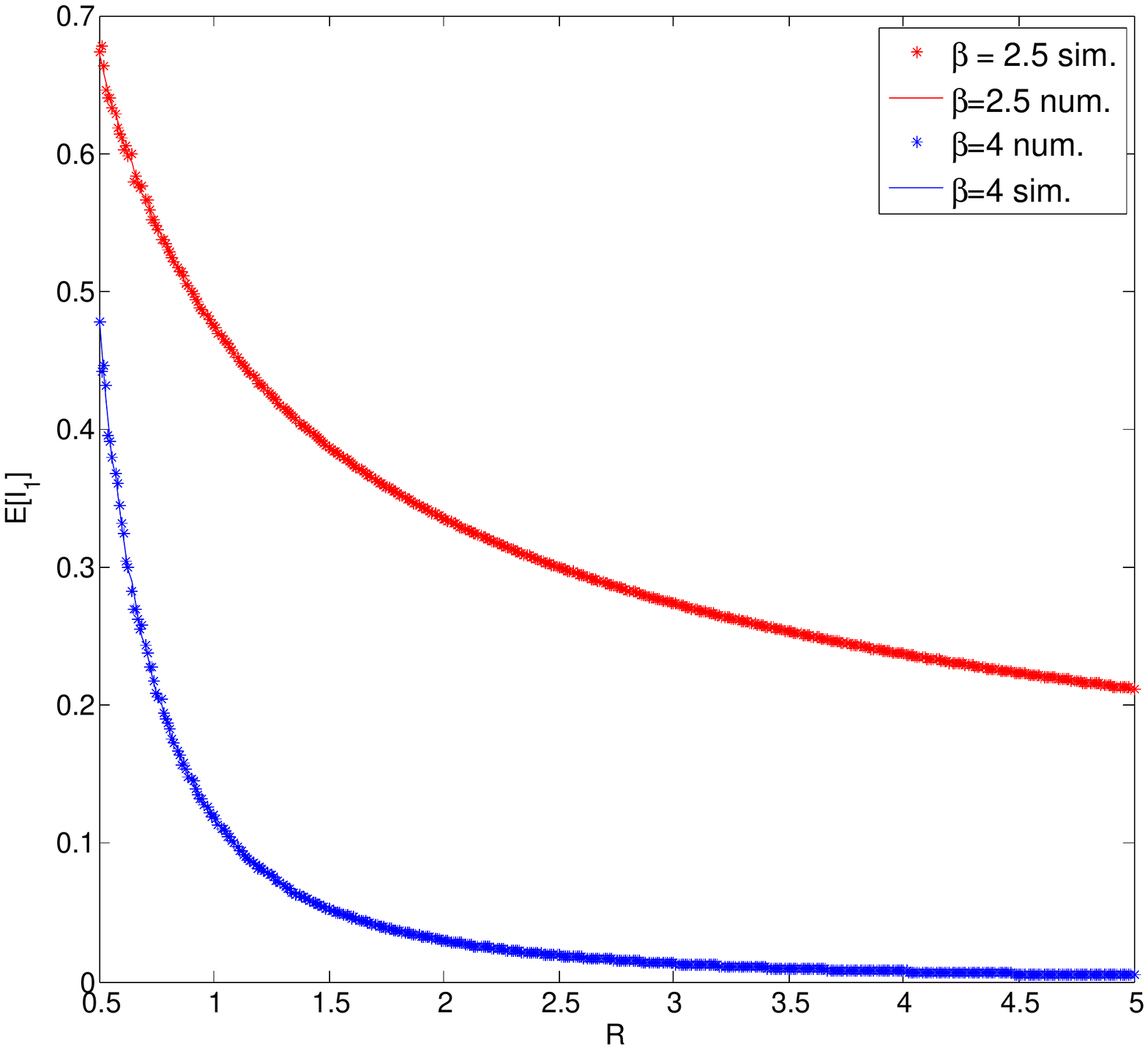}
       	\caption{Interference generated by the single atoms outside a ball of radius R. The upper plot shows the case with path-loss exponent $\beta=2.5$ and the lower plot the case with $\beta=4$.}
	\label{SingleIR}
\end{figure}

Specifically for $\mathcal{I}_1$, we first transform in polar coordinates (\ref{EPhi1}) and then perform numerical integration, for the function $f(r_x)=h(0,r_x)\mathbbm{1}_{\left\{r_x> R\right\}}$, with $\nu(0,(r_x,\theta_x)):=1$. Here, $R$ is a positive distance, within the interval $R\in\left[0.5,5\right]$. The indicator function is used in order to calculate the interference created by singles outside a ball centred at $0$ and of radius $R$. The evaluation is shown in Fig. \ref{SingleIR} with a continuous line. The results from the simulations are shown by the star-dotted curve. With the appropriate window, the expression in (\ref{Expected}) gives almost identical results with the simulations. 

\begin{figure}[t!]
	\centering  
	\includegraphics[trim = 0mm 50mm 5mm 60mm, clip, width=0.35\textwidth]{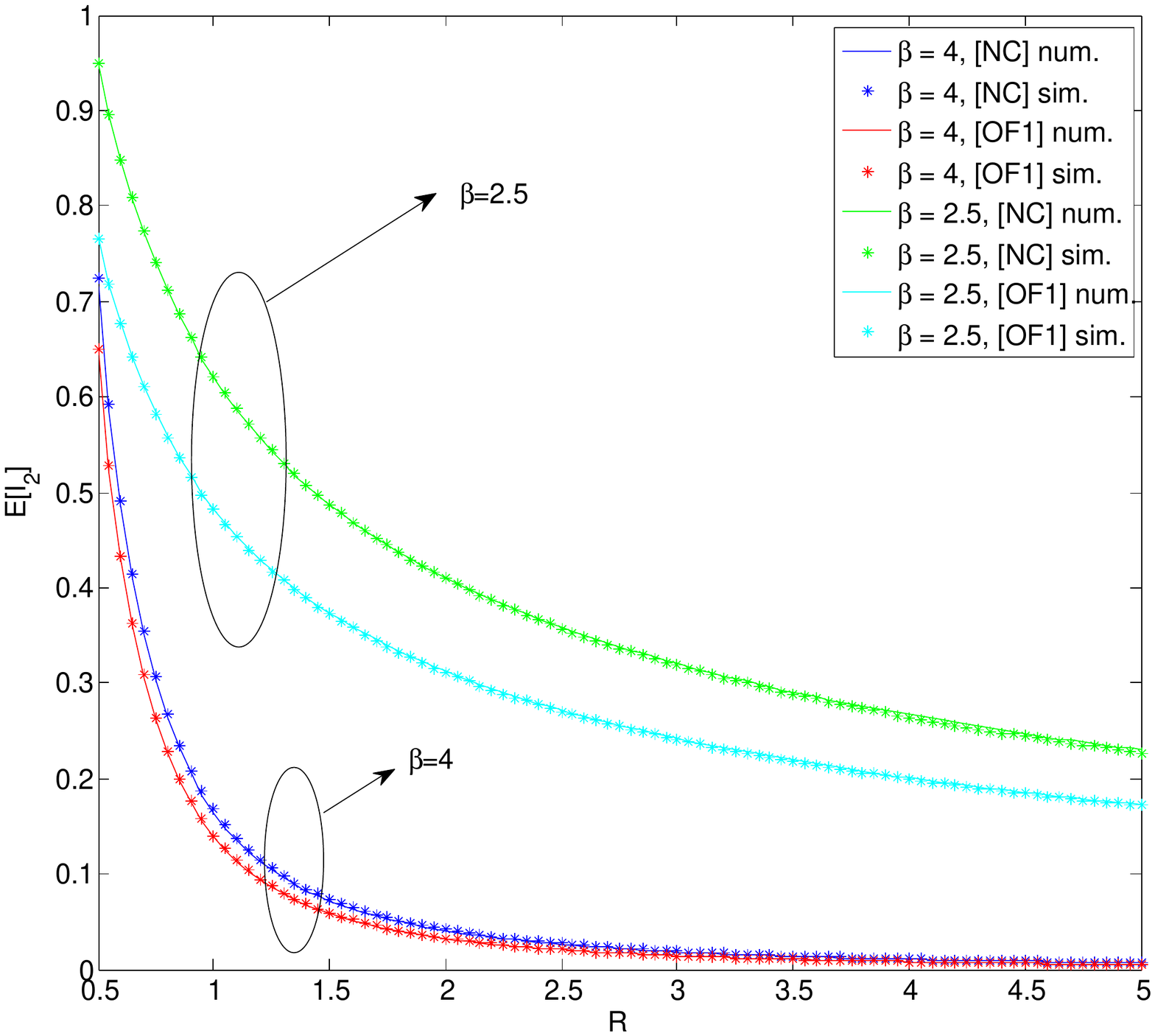}
       	\caption{Interference generated by pairs of atoms outside a ball of radius R. The upper two plot shows the case with path-loss exponent $\beta=2.5$ and the lower two plots the case with $\beta=4$. We show the two cooperation scenarios [NC] and [OF1].}
	\label{DoubleIR}
\end{figure}

Similarly for $\mathcal{I}_2$, numerical evaluation and simulation results using $g(r_x,r_y)\mathbbm{1}_{\left\{r_x,r_y> R\right\}}$ are shown in Fig.\ref{DoubleIR}. For the numerical integration, we evaluate the two cases [NC] and [OF1] for different path-loss exponent. Obviously interference from [OF1] is always less than [NC] since it is received only from one of the two BSs of each pair, while the other is silent. Nevertheless, the two scenarios do not numerically defer much for $\beta=4$, as shown in the figure.

\section{General case and Conclusions}
The analysis for the dependent thinning of a PPP using the NNM can be generalised to include groups of atoms with size greater than two. For the computation of such probabilities, certain difficulties are raised due to the overlapping of more than two discs with different radii. However, the methodology remains the same. Results on the percentage of atoms, Voronoi surface and repulsion/attraction, which are independent of the PPP density, can be derived by Monte Carlo simulations. The same analysis for the expectation and the LT of the interference can also be followed in cases of larger groups. 

Altogether, in this paper we proposed a method to define static groups of cooperating Base Stations based on the Nearest Neighbour Model and analyse them with the use of stochastic geometry. Many structural characteristics were derived for the case of singles and pairs. These can be further used in the analysis of $\mathrm{SINR}$ models and their performance evaluation. We gave some first results by calculating the expected value and the Laplace transform of the interference from different groups. Further analysis and applications in this direction is the current research work of the authors. 

\section*{Acknowledgment}
The authors gratefully thank Prof. Fran\c{c}ois Baccelli for suggesting the Nearest Neighbour Model and providing related literature. 

\section*{Appendix}
\label{Appendix}
\subsection*{Proof of Theorem \ref{Percentage}.}
\label{AppendixA}

Let us condition on the fact that an atom lies at the planar point $x$ and then find the probability that this is in pair (single). This is written as $\mathbb{P}^x\left( x \overset{\Phi}{\leftrightarrow} y, \mbox{ for some } y\in  \Phi\setminus{\left\{x\right\}} \right)$, where $\mathbb{P}^x$ is the Palm measure. We use Slivnyak-Mecke's theorem \cite{BacBlaVol1} , which states that $\Phi\setminus\left\{x\right\}$ under $\mathbb{P}^{x}$ has the same distribution as $\Phi$ under $\mathbb{P}$.

\begin{equation*}
\begin{split}
\mathbb{P}\left( x \overset{\Phi}{\leftrightarrow} y, \mbox{ for some } y\in  \Phi \right)= & \mathbb{E}\left( \mathbbm{1}_{\left\{ x \overset{\Phi}{\leftrightarrow} y, \mbox{ for some } y\in  \Phi \right\}} \right)\\
\stackrel{(a)}{=} & \mathbb{E}\left( \sum_{y\in \Phi} \mathbbm{1}_{ \left\{ x \overset{\Phi}{\leftrightarrow} y \right\}} \right)\\
\stackrel{(b)}{=} & \int_{\mathbb{R}^2} \mathbb{E} \left( \mathbbm{1}_{\{ x \overset{\Phi}{\leftrightarrow} y\}}\right) \lambda dy\\
= & \int_{\mathbb{R}^2} \mathbb{P} \left( x \overset{\Phi}{\leftrightarrow} y \right)\lambda dy\\
\stackrel{(c)}{=}  & \int_{\mathbb{R}^2} e^{-\lambda\pi |x-y|^2(2-\gamma)} \lambda dy\\
= & \int^\infty_0 \int^{2 \pi}_0 e^{-\lambda \pi r^2(2-\gamma)} \lambda d\theta rdr \\
= &  \frac{1}{(2-\gamma)} \approx \ 	 0.6215.
\end{split}
\end{equation*} 
Equality (a) holds because for PPPs the nearest neighbour of an atom is a.s. unique, so we cannot have three atoms $x,y,w\in\phi$ with $y\neq w$, such that $x \overset{\phi}{\leftrightarrow} y$ and $x \overset{\phi}{\leftrightarrow} w$. Equality (b) comes from Campbell's formula \cite{BacBlaVol1} and in (c) we use Lemma \ref{Lemma1}.

Similarly, to find the probability of $x\in\Phi$ being single
\begin{equation*}
\begin{split}
\mathbb{P}^{x}\Big( x^{\Phi}_{\#} \Big)= & 1-\mathbb{P}^{x}\left( x \overset{\Phi}{\leftrightarrow} y, \mbox{ for some } y\setminus{\left\{x\right\}}\in  \Phi \right) \\
= &	 \frac{1-\gamma}{2-\gamma}\approx 0.3785.
\end{split}
\end{equation*}

\subsection*{Palm measures and the \textit{NN} function of $\Phi^{(2)}$.}
\label{AppendixB}

Let $\left(\Omega,\mathcal{F},\mathbb{P}\right)$ be the underlying probability space and $\Phi$ a PPP. The Palm measure of $\Phi$ is $\mathbb{P}^x$. We denote by $\mathbb{P}^{(1),x}$ and $\mathbb{P}^{(2),x}$ the Palm measure of $\Phi^{(1)}$ and $\Phi^{(2)}$, respectively. Let us first consider $\mathbb{P}^{(2),x}$ and use the heuristic definition $\mathbb{P}^{(2),x}(\cdot)=\mathbb{P}(\cdot | x\in \Phi^{(2)})$, where we condition on the existence of one atom of $\Phi$ in a small neighbourhood of $x$. By Definition \ref{defi1}, $x\in \phi^{(2)}$ if, and only if, $\mathcal{A}_x$ holds, where
\begin{equation*}
\begin{split}
\mathcal{A}_x & =\big\{\mbox{there exists some } y\in \phi\backslash \{x\} \mbox{ such that } x\leftrightarrow y \big\}\\
& = \left\{\sum_{y\in \phi\backslash \{x\}}\mathbbm{1}_{\{x \overset{\phi}{\leftrightarrow} y\}}= 1\right\}.
\end{split}
\end{equation*}
Also, from Theorem \ref{Percentage}, $\mathbb{P}^x(\mathcal{A}_x)=p^*$ for PPPs. For every $\Gamma \in \mathcal{F}$,
\begin{eqnarray}
\label{P2x}
\begin{split}
\mathbb{P}^{(2),x}(\Gamma) & =\mathbb{P}(\ \Gamma \ | x\in \Phi,\mathcal{A}_x)\\
& =\frac{\mathbb{P}(\Gamma , \mathcal{A}_x | x\in \Phi)}{\mathbb{P}(\mathcal{A}_x|x\in \Phi)}\\
& =\frac{\mathbb{P}^x(\Gamma , \mathcal{A}_x )}{\mathbb{P}^x(\mathcal{A}_x)}\\
& =\mathbb{P}^x(\Gamma | \mathcal{A}_x ) \\
& =\mathbb{P}^x(\Gamma , \mathcal{A}_x )\frac{1}{p^*}.
\end{split}
\end{eqnarray}
For a rigorous proof of the Palm measure for $\Phi^{(2)}$, we refer the reader to the last part of the Appendix. 

To calculate the \textit{NN function} of $\Phi^{(2)}$, let us take $r\geq 0$ and consider $\Gamma=\big\{d(x,\Phi^{(2)}\backslash \{x\})\leq r\big\}$. Then,
\begin{equation*}
\begin{split}
G^{(2)}(r) & =\mathbb{P}^{(2),x}\Big(d(x,\Phi^{(2)}\backslash \{x\})\leq r\Big)\\
& \stackrel{(\ref{P2x})}{=} \mathbb{P}^x\Big(d(x,\Phi^{(2)}\backslash \{x\})\leq r , \mathcal{A}_x \Big)\frac{1}{p^*}\\
& =\mathbb{E}^x\left( \sum_{y\in \Phi\backslash \{x\}}\mathbbm{1}_{\{ d(x,\Phi^{(2)}\backslash \{x\})\leq r , x\leftrightarrow y \} }\right)\frac{1}{p^*}.\\
\end{split}
\end{equation*}
For every $\phi$ and $y\in \phi\backslash \{x\}$, if $x\leftrightarrow y$, it is true that 
\begin{equation}
\label{equality}
d(x,\phi^{(2)}\backslash \{x\})=d(x,\phi \backslash \{x\})=d(x,y).
\end{equation}
We use this observation, Slivnyak-Mecke's theorem, the Campbell-Little-Mecke formula, and Lemma \ref{Lemma1}, to find that
\begin{equation*}
\begin{split}
G^{(2)}(r) & =\mathbb{E}\left( \sum_{y\in \Phi}\mathbbm{1}_{\{ d(x,y)\leq r , x\leftrightarrow y \} }\right)\frac{1}{p^*}\\
& =\mathbb{E}\left( \sum_{y\in \Phi}\mathbbm{1}_{\{ d(0,y)\leq r , 0\leftrightarrow y \} }\right)\frac{1}{p^*}\\
& =\int_{\mathbb{R}^2}\mathbb{E}\left( \mathbbm{1}_{\{ d(0,y)\leq r , 0\leftrightarrow y \} }\right)\lambda dy \frac{1}{p^*}\\
& =\int_{\mathbb{R}^2}\mathbb{P}\Big( d(0,y)\leq r , 0\leftrightarrow y \Big)\lambda dy \frac{1}{p^*}\\
& =\int^{2\pi}_0 \int^\infty_0\mathbb{P}\Big( s\leq r , 0\leftrightarrow (s,\theta) \Big)\lambda s ds d\theta \frac{1}{p^*}\\
& =2\pi \lambda \int^r_0\mathbb{P}\Big( 0\leftrightarrow (s,\theta) \Big) s ds \frac{1}{p^*}\\
& =2\pi \lambda \int^r_0e^{-\lambda \pi s^2(2-\gamma)} s ds \frac{1}{p^*}\\
&=1-e^{-\lambda \pi r^2(2-\gamma)},\\
\end{split}
\end{equation*}
where the last equality is due to the fact that $p^*=\frac{1}{(2-\gamma)}$.

We further remind the reader that $x\in \phi^{(1)}$ if, and only if, the event $\mathcal{B}_x$ holds, where 
\begin{equation*}
\mathcal{B}_x=\{\mbox{ for every } y\in \phi\backslash \{x\} \mbox{ such that } x\rightarrow y, \ y\nrightarrow x\}.
\end{equation*}
It is possible to give a similar expression to $\mathbb{P}^{(1),x}$ in the same way as we did for $\mathbb{P}^{(2),x}$. For every $\Gamma \in \mathcal{F}$,
\begin{eqnarray*}
\mathbb{P}^{(1),x}(\Gamma) & =\mathbb{P}^x(\Gamma | \mathcal{B}_x) & = \frac{\mathbb{P}^x(\Gamma , \mathcal{B}_x)}{(1-p^*)}.
\end{eqnarray*}
It has not been possible however to make a similar analysis to get the \textit{NN function} of $\Phi^{(1)}$. The reason is that it is not easy to precisely define the nearest neighbour atom in this case. A property similar to that in equation \eqref{equality}, which was crucial in the previous proof, could not be found.

\subsection*{Proof of Theorem \ref{Expected}.}
\label{AppendixC}

Let us start with $\mathcal{I}^{(1)}$. We observe that 
\begin{equation*}
\mathcal{I}^{(1)} = \sum_{x\in \phi^{(1)}}f(x)=\sum_{x\in \phi}f(x)\mathbbm{1}_{\left\{x^\Phi_{\#}\right\}}\ \mathbb{P}\ a.s.,
\end{equation*}
and that, for every $x\in\phi$
\begin{equation*}
\begin{split}
\mathbbm{1}_{\{x^\phi_{\#}\}} = & \sum_{y\in \phi \backslash \{x\}}\mathbbm{1}_{\left\{x\overset{\phi}{\rightarrow} y,y \overset{\phi}{\nrightarrow} x \right\}}
\end{split}
\end{equation*}
By the reduced Campbell-Little-Mecke formula and Slivnyak-Mecke's Theorem,
\begin{equation*}
\begin{split}
\mathbb{E}\left[\mathcal{I}^{(1)} \right]= & \mathbb{E}\left[\sum_{x\in \Phi}\sum_{y\in \Phi \backslash \{x\} }f(x)\mathbbm{1}_{\left\{x\overset{\phi}{\rightarrow} y,y \overset{\phi}{\nrightarrow} x \right\}}\right]\\
= & \int_{\mathbb{R}^2}\int_{\mathbb{R}^2}\mathbb{E}\left[f(x)\mathbbm{1}_{\left\{x\overset{\phi}{\rightarrow} y,y \overset{\phi}{\nrightarrow} x \right\}}\right]\lambda dy \lambda dx\\
= & \int_{\mathbb{R}^2}\mathbb{E}\left[f(x)\right]\int_{\mathbb{R}^2}\mathbb{E}\left[\mathbbm{1}_{\left\{x\overset{\phi}{\rightarrow} y,y \overset{\phi}{\nrightarrow} x \right\}}\right]\lambda dy \lambda dx\\
\stackrel{(a)}{=} & \int_{\mathbb{R}^2}\mathbb{E}\left[f(x)\right]\mathbb{P}\big(x^\Phi_{\#}\big) \lambda dx\\
\stackrel{(b)}{=}  & (1-p^*) \int_{\mathbb{R}^2}\mathbb{E}\left[f(x)\right] \lambda dx,\\
\end{split}
\end{equation*}
where (a) and (b) come from the proof of Theorem \ref{Percentage}.

For $\mathcal{I}^{(2)}$, we make the observation that
\begin{equation*}
\begin{split}
& \sum_{x\in \phi^{(2)}} \sum_{y\in \phi^{(2)}}^{\neq}g(x,y)\mathbbm{1}_{\left\{x \overset{\phi^{(2)}}{\leftrightarrow} y\right\}} = \sum_{x\in \phi} \sum_{y\in \phi}^{\neq} g(x,y)\mathbbm{1}_{\left\{x \overset{\phi}{\leftrightarrow} y\right\}}.
\end{split}
\end{equation*}
Then, as previously (and using Theorem \ref{Percentage}) we can calculate the expected value of the interference from pairs
\begin{equation*}
\begin{split}
\mathbb{E}\left[\mathcal{I}^{(2)}\right]= & \mathbb{E}\left[\sum_{x\in \Phi} \sum_{y\in \Phi\setminus\left\{x\right\}} g(x,y)\mathbbm{1}_{\left\{x \overset{\Phi}{\leftrightarrow} y\right\}}\right]\\
\stackrel{(c)}{=} & \int_{\mathbb{R}^2}\int_{\mathbb{R}^2} \mathbb{E}\left[g(x,y)\mathbbm{1}_{\left\{x \overset{\Phi}{\leftrightarrow} y\right\}}\right] \lambda dy \lambda dx \\
= & \int_{\mathbb{R}^2}\int_{\mathbb{R}^2} \mathbb{E}\left[g(x,y)\right]\mathbb{P}\left(x \overset{\Phi}{\leftrightarrow} y\right) \lambda dy \lambda dx\\
\stackrel{(d)}{=} & \int_{\mathbb{R}^2}\int_{\mathbb{R}^2} \mathbb{E}\left[g(x,y)\right] e^{-\lambda\pi|x-y|^2(2-\gamma)}\lambda dy \lambda dx,\\
\end{split}
\end{equation*}
(c) uses Campbell's formula and (d) comes from Lemma \ref{Lemma1}.

\subsection*{Proof of Corollary \ref{CorIM}.}
Let us take $A\in \mathcal{B} \left(\mathbb{R}^2\right)$. We use in (\ref{I1}) the function $f(x)=\mathbbm{1}_{\left\{x\in A\right\}}$, which indicates whether the atom $x$ belongs to $A$ or not. We thus have


\begin{equation*}
\begin{split}
\mathcal{I}^{(1)} & =\sum_{x\in \phi^{(1)}}f(x) \\
& = \sum_{x\in \phi^{(1)}}\mathbbm{1}_{\left\{x\in A\right\}},
\end{split}
\end{equation*}
which counts the number of elements of $\phi^{(1)}$ inside $A$, and, hence, its expected value is $\Phi^{(1)}$'s intensity measure

\begin{equation*}
\begin{split}
\mathbb{E}\left[ \sum_{x\in \Phi^{(1)}}\mathbbm{1}_{\left\{x\in A\right\}}\right] = \mathbb{E} \left[ \Phi^{(1)}(A) \right] = M^{(1)}(A).
\end{split}
\end{equation*}
The righthand side in (\ref{EPhi1}) of Theorem \ref{Expected} is 

\begin{equation*}
\begin{split}
\int_{\mathbb{R}^2}f(x) (1-p^*) \lambda dx & =  \int_A (1-p^*) \lambda dx, \\
\end{split}
\end{equation*}
and we conclude that 

\begin{equation*}
M^{(1)}(dx)=(1-p^*)\lambda dx.
\end{equation*} 

For $\Phi^{(2)}$ we consider again the function $g(x,y)=\mathbbm{1}_{\left\{x\in A\right\}}$. Given that, for every $x\in \phi^{(2)}$, $\sum_{y\in \phi^{(2)}\backslash \{x\}} \mathbbm{1}_{\{ x \stackrel{\phi^{(2)}}{\leftrightarrow} y\}} = 1 $, equation (\ref{I2}) takes the form

\begin{equation*}
\begin{split}
\mathcal{I}^{(2)} 
& = \sum_{x\in \phi^{(2)}} \mathbbm{1}_{\left\{x\in A\right\}}\sum_{y\in \phi^{(2)}\backslash \{x\}} \mathbbm{1}_{\{ x \stackrel{\phi^{(2)}}{\leftrightarrow} y\}} \\
& = \sum_{x\in \phi^{(2)}} \mathbbm{1}_{\left\{x\in A\right\}}.
\end{split}
\end{equation*}
Its expected value is $\Phi^{(2)}$'s intensity measure

\begin{equation*}
\begin{split}
\mathbb{E}\left[ \sum_{x\in \phi^{(2)}} \mathbbm{1}_{\left\{x\in A\right\}}\right] & = \mathbb{E} \left[\Phi^{(2)}(A) \right] = M^{(2)}(A).
\end{split}
\end{equation*}
The righthand side in (\ref{EPhi2}) of Theorem \ref{Expected} is 

\begin{eqnarray*}
\begin{split}
& \int_{\mathbb{R}^2} \int_{\mathbb{R}^2} g(x,y)e^{-\lambda \pi |x-y|^2(2-\gamma)}\lambda dy \lambda dx =\nonumber\\ 
& \int_{A} \int_{\mathbb{R}^2} \mathbbm{1}^{(x)}_Ae^{-\lambda \pi |x-y|^2(2-\gamma)}\lambda dy \lambda dx = \nonumber\\
& \int_{A} \int_{\mathbb{R}^2} e^{-\lambda \pi |x-y|^2(2-\gamma)}\lambda dy \lambda dx = \int_{A} p^* \lambda dx. \nonumber
\end{split}
\end{eqnarray*}
We conclude that 

\begin{equation*}
M^{(2)}(dx)=p^*\lambda dx.
\end{equation*} 
This result provides an analytical argument of the stationarity of the processes $\Phi^{(1)}$ and $\Phi^{(2)}$ through an explicit formula of their intensity measure and their intensity coefficient. Moreover, using the intensity measure. we can find the average number of points of $\Phi^{(1)}$ or $\Phi^{(2)}$ over any Borel set. This clarifies our discussion at the end of section III.A under Theorem \ref{Percentage}.


\subsection*{Palm Measure for $\Phi^{(2)}$ - rigorous proof}

In this section, $\Phi$ is a stationary point process, with intensity $\lambda>0$, and Palm measure given by $\mathbb{P}^0$. For a realisation $\phi=\{x_n\}$ and a fixed $x\in \mathbb{R}^2$, we denote by $\phi_{x}:=\{x_n+x\}$, the translation of $\phi$ by $x$.

In the previous sections, we have defined  the processes $\Phi^{(1)}$ and $\Phi^{(1)}$ only by geometric means and not making use of the probability law that governs $\Phi$. What we have said is that given a realisation $\phi$, 
\begin{equation*}
\begin{split}
&\phi^{(1)}=\{x\in \phi \ | \ x \mbox{ is single }\},\\
&\phi^{(2)}=\{x\in \phi \ | \ x \mbox{ cooperates with another element of } \phi\}.
\end{split}
\end{equation*}
Denoting by $\mathbb{M}$ the space of realisations, it is possible to consider
\begin{eqnarray*}
\begin{split}
& \phi \mapsto \phi^{(1)} & & \phi  \mapsto \phi^{(2)}, 
\end{split}
\end{eqnarray*}
as measurable mappings $ \mathbb{M} \rightarrow \mathbb{M}$. Since we use mappings over the space of realisations, we can not say much about the probability laws governing $\Phi^{(1)}$ and $\Phi^{(2)}$, which we denote by $P^{(1)}$ and $P^{(2)}$, respectively. Let us consider the event $\Gamma\in\mathcal{F}$. Each $P^{(i)}$ is a measure 
\begin{equation}
\begin{split}
P^{(i)}: & \mathcal{F}\rightarrow [0,1] \\
         & \Gamma \rightarrow P^{(i)}(\Gamma)=\mathbb{P}(\Phi^{(i)}\in \Gamma).
\end{split}
\end{equation}
We define $\Gamma^{(i)}\in \mathcal{F}$ as $\Gamma^{(i)}:=\{\phi \ | \ \phi^{(i)}\in \Gamma \}$, therefore, 
\begin{equation}
P^{(i)}(\Gamma)=\mathbb{P}\left(\Phi \in \Gamma^{(i)}\right)=\mathbb{E}\left(\mathbbm{1}_{\{\Phi\in \Gamma^{(i)} \}} \right).
\end{equation}
(Let us note here that this last equation gives us a simple way to approximate $P^{(i)}$ using the Monte Carlo method.)

Since $\Phi$ is stationary, the Palm probability of $\Phi$ is defined as follows \cite[p.119]{StoKMbook95}, \cite[p.51,(24)]{BaddNotes07}. For every $A\in \mathcal{B}(\mathbb{R}^2)$, and every $\hat{\Gamma} \in \mathcal{F}$, 
\begin{equation}\label{PalmOr}
\mathbb{E}\sum_{x\in \Phi} \mathbbm{1}_{\{x\in A\}}\mathbbm{1}_{\left\{\Phi_{-x}\in \hat{\Gamma} \right\}}= \lambda \mathcal{S}(A)\mathbb{P}^{0}\left(\hat{\Gamma}\right).
\end{equation}

Using the above definition, we want to find a probability measure $P^{(2),0}:\mathcal{F}\rightarrow [0,1]$ such that, for every $A\in \mathcal{B}(\mathbb{R}^2)$, and every $\Gamma \in \mathcal{F}$, 
\begin{equation}\label{PalmPr}
\mathbb{E}\sum_{x\in \Phi^{(2)}} \mathbbm{1}_{\{x\in A\}}\mathbbm{1}_{\{\Phi^{(2)}_{-x}\in \Gamma\}}= p^* \lambda \mathcal{S}(A)P^{(2),0}(\Gamma).
\end{equation} 


(Observe that we use the different events $\Gamma$ and $\hat{\Gamma}$ for reasons that will be clear later in the proof). Let us denote by $\mathcal{A}_0:=\{\phi \ | \mbox{ there exists } y\in \phi \mbox{ such that } 0 \stackrel{\phi}{\leftrightarrow} y \}$. For a given realisation $\phi$, it holds $x\in \phi^{(2)}$ if, and only if, $\phi_{-x} \in \mathcal{A}_0$. 

We assume that the Palm measure for $\Phi^{(2)}$ is $P^{(2),0}(\Gamma):=\mathbb{P}^{0}\left( \Gamma^{(2)}\big| \mathcal{A}_0 \right)$ and we want to verify that the measure satisfies (\ref{PalmPr}). Since, $\mathbb{P}^{0}\left(\mathcal{A}_0 \right)=p^*>0$,  it is definitely a probability measure. Let us take $A\in \mathcal{B}(\mathbb{R}^2)$ and $\Gamma \in \mathcal{F}$. We start by the righthand side of (\ref{PalmPr})
\begin{equation*}
\begin{split}
p^*\lambda \mathcal{S}(A)P^{(2),0}(\Gamma) & = p^*\lambda \mathcal{S}(A)\frac{\mathbb{P}^{0}\left( \Gamma^{(2)}, \mathcal{A}_0 \right)}{p^*}\\
& = \lambda \mathcal{S}(A)\mathbb{P}^{0}\left( \Gamma^{(2)}, \mathcal{A}_0 \right)\\
& \stackrel{(a)}{=} \lambda \mathcal{S}(A)\mathbb{P}^{0}\left( \hat{\Gamma} \right)\\
& \stackrel{(\ref{PalmOr})}{=} \mathbb{E}\sum_{x\in \Phi} \mathbbm{1}_{\{x\in A\}}\mathbbm{1}_{\left\{\Phi_{-x}\in \hat{\Gamma} \right\}}\\
& = \mathbb{E}\sum_{x\in \Phi} \mathbbm{1}_{\{x\in A\}}\mathbbm{1}_{\left\{\Phi_{-x}\in \Gamma^{(2)},\mathcal{A}_0 \right\}}\\
& = \mathbb{E}\sum_{x\in \Phi} \mathbbm{1}_{\{x\in A\}}\mathbbm{1}_{\left\{\Phi_{-x}\in \Gamma^{(2)} \right\}}\mathbbm{1}_{\left\{\Phi_{-x}\in\mathcal{A}_0 \right\}} \\ 
& \stackrel{(b)}{=} \mathbb{E}\sum_{x\in \Phi^{(2)}} \mathbbm{1}_{\{x\in A\}}\mathbbm{1}_{\left\{\Phi_{-x}\in \Gamma^{(2)} \right\}}\\
& \stackrel{(c)}{=} \mathbb{E}\sum_{x\in \Phi^{(2)}} \mathbbm{1}_{\{x\in A\}}\mathbbm{1}_{\left\{ \Phi^{(2)}_{-x} \in \Gamma \right\}}. \\  
\end{split}
\end{equation*}
So we reached the lefthand side of (\ref{PalmPr}). In the above, (a) comes by replacing $\hat{\Gamma}:= \left\{\Gamma^{(2)},\mathcal{A}_0\right\}$, (b) because $x\in \phi^{(2)}$ if, and only if $\phi_{-x} \in \mathcal{A}_0$, and (c) because by definition $\phi_{-x}\in\Gamma^{(2)}$ if, and only if $\phi^{(2)}_{-x} \in \Gamma$.

In a similar way, for $P^{(1),0}:\mathcal{F}\rightarrow [0,1]$, and for every $\Gamma\in\mathcal{F}$ we can show, using the same arguments, that $P^{(1),0}(\Gamma):=\mathbb{P}^{0}\left( \Gamma^{(1)}\big| \ \mathcal{B}_0 \right)$  is a Palm measure for $\Phi^{(1)}$.

\bibliographystyle{unsrt}

%


\end{document}